\documentclass[fleqn,usenatbib]{mnras}

\usepackage{mathptmx}
\usepackage[T1]{fontenc}
\usepackage{ulem}
\usepackage{color}

\DeclareRobustCommand{\VAN}[3]{#2}
\let\VANthebibliography\thebibliography
\def\thebibliography{\DeclareRobustCommand{\VAN}[3]{##3}\VANthebibliography}

\usepackage{graphicx}	 % Including figure files
\usepackage{amsmath}	 % Advanced maths commands
\usepackage{amssymb}	 % Extra maths symbols
\usepackage[switch]{lineno}

\title[Magnetic-reconnection-heated corona in luminous AGNs]{Magnetic-Reconnection-Heated Corona Model: Implication of Hybrid Electrons for Hard X-ray Emission of Luminous Active Galactic Nuclei}

\author[J.Y. Liu et al.]{
Jie-Ying Liu,$^{1,2,3,4}$
Jirong Mao,$^{1,2,3}$\thanks{E-mail: jirongmao@mail.ynao.ac.cn}
B. F. Liu$^{5,6}$\thanks{E-mail: bfliu@nao.cas.cn}
\\
 $^{1}$Yunnan Observatories, Chinese Academy of Sciences, Kunming 650216, Yunnan Province, China \\
$^{2}$Center for Astronomical Mega-Science, Chinese Academy of Sciences, 20A Datun Road, Chaoyang District, Beijing 100012, China \\
 $^{3}$Key Laboratory for the Structure and Evolution of Celestial Objects, Chinese Academy of Sciences, Kunming 650216, China\\
$^{4}$International Centre of Supernovae, Yunnan Key Laboratory, Kunming 650216, P. R. China\\
$^{5}$Key Laboratory of Space Astronomy and Technology, National Astronomical
Observatories, Chinese Academy of Sciences, Beijing 100012,China\\
$^{6}$School of Astronomy and Space Sciences, University of
Chinese Academy of Sciences, 19A Yuquan Road, Beijing 100049, China
}
\date{Accepted 2023 November 17. Received 2023 November 14; in original form 2023 March 8}

\pubyear{2023}

\begin{document}

\label{firstpage}
\pagerange{\pageref{firstpage}--\pageref{lastpage}}
\maketitle

% Abstract of the paper
\begin{abstract}
It is widely accepted that X-ray emission in luminous active galactic nuclei (AGNs) originates from hot corona. To prevent the corona from over-cooling by strong X-ray emission, steady heating to the corona is essential, for which the most promising mechanisms is the magnetic reconnection. Detailed studies of the coupled disc and corona, in the frame of magnetic field transferring accretion-released energy from the disc to the corona, reveal that the thermal electrons can only produce X-ray spectrum with $\Gamma_{\rm 2-10\,keV}>2.1$, which is an inevitable consequence of the radiative coupling of the thermal corona and disc. In the present work, we develop the magnetic-reconnection-heated corona model by taking into account the potential non-thermal electrons accelerated in the magnetic reconnection process, in addition to the thermal electrons. We show that the features of the structure and spectrum of the coupled disc and corona can be affected by the fraction of magnetic energy allocated to thermal electrons. Furthermore, we investigate the effects of the power-law index and energy range of non-thermal electrons and the magnetic field on the spectrum. It is found that the X-ray spectrum from the Comptonization of the hybrid electrons can be flatter than that from thermal electrons only, in agreement with observations. By comparing with the observed hard X-ray data, we suggest that a large fraction ($>40\%$) of the magnetic energy be allocated to the non-thermal electrons in the luminous and flat X-ray spectrum AGNs.
\end{abstract}

% Select between one and six entries from the list of approved keywords.
% Don't make up new ones.

%%%%%%%%%%%%%%%%%%%%%%%%%%%%%%%%%%%%%%%%%%%%%%%%%%

%%%%%%%%%%%%%%%%% BODY OF PAPER %%%%%%%%%%%%%%%%%%
 
%\pagewiselinenumbers %
\section{Introduction}\label{sec:intro}
It is believed that active galactic nuclei (AGNs) are powered by the gravitational energy liberation of the accretion matter. Luminous AGNs generally have relatively high Eddington ratio of $L_{\rm bol}/L_{\rm Edd}>0.02 $, where $L_{\rm bol}$ is the bolometric luminosity, and $L_{\rm Edd}=1.25\times10^{38}(M_{\rm BH}/M_{\odot})$\,erg\,s$^{-1}$ is the Eddington luminosity. Their spectral features are characterized by three components: the so-called ``big blue bump'' ranging from the optical to the ultraviolet (UV) band, which is explained by a standard thin disc extending into the innermost stable circular orbit \citep[e.g.,][]{1973A&A....24..337S}, the soft X-ray excess, whose origin is under debate \citep[e.g.,][where various origins were summarized]{2007ASPC..373..121D}, and the power-law emission in hard X-ray band that is commonly contributed by the inverse Compton scattering of optical/UV photons from the accretion disc. Because the low-energy photons are scattered up to X-ray band through the interaction with hot electrons in the corona \citep{1994ApJ...436..599S,1998MNRAS.301..179M,2000ApJ...542..703Z,2016MNRAS.458.2454L}, the observational feature of the luminous AGNs in X-ray band is the key point to reveal the relation between the disc and the corona at different accretion rates \citep[]{2004ApJ...607L.107W, 2006ApJ...646L..29S, 2007MNRAS.381.1235V, 2010ApJ...720L.206Z, 2019A&A...628A.135A,2019MNRAS.487.3884C, 2020MNRAS.491.2576W}.

Heating mechanism for AGN corona against strong inverse Compton cooling is unclear. Magnetic energy is suggested to be a major energy source that is carried into the corona by magnetic field and released as thermal energy through the magnetic reconnection process \citep[e.g.,][]{1998MNRAS.299L..15D,2000ApJ...534..398M,2001MNRAS.321..549M,2002ApJ...572L.173L,
2003ApJ...587..571L,2003MNRAS.345.1057M,2004ApJ...607L.107W}. In particular,
\cite{2003ApJ...587..571L} obtained both soft and hard X-ray spectra in different AGN
accretion cases by the disc-corona model, where the corona is heated by the magnetic reconnection.
Applying the magnetic-reconnection-heated corona model, one can well explain the positive correlation between the hard X-ray bolometric correction factor and the Eddington ratio. Meanwhile, the trend of the X-ray photon index increasing with the Eddington ratio can be also interpreted \citep{2009MNRAS.394..207C, 2012ApJ...761..109Y, 2016ApJ...833...35L}.
 The coupling of the magnetic-reconnection-heated disc-corona and the jet was used to illustrate the ``outlier'' relation of $L_{\rm R}\propto L_{\rm X}^{\sim1.4}$ in the black hole X-ray binaries by \cite{2015MNRAS.448.1099Q}. They suggested that the X-ray emission of $2$--$10$\,keV is dominated by the radiation of the disc-corona system. In order to reproduce the observed data of optical/UV and X-ray band in luminous AGNs, \cite{2020MNRAS.495.1158C} further refined the model in \cite{2016ApJ...833...35L}. They set the outer boundary of the disc-corona accretion flow as the self-gravity radius and efficiently considered both local and global energy-conservation. They found that the accretion energy transported to the corona depends on both the magnetic field and the accretion rate, which is consistent with the results found by \cite{2016ApJ...833...35L}. They successfully applied this model to fit the spectra of 16 luminous AGNs.

We notice that the former magnetic-reconnection-heated corona model reproduces a relatively steep X-ray spectrum with the photon index $\Gamma_{\rm 2-10\,keV}>2.1$ \citep{2003ApJ...587..571L, 2009MNRAS.394..207C, 2016ApJ...833...35L, 2020MNRAS.495.1158C}. Such steep spectra are understood as a consequence of the disc-corona coupling. When soft photons from the disc are up-scattered in the corona, it causes cooling of electrons and condensation of coronal gas to the disc. The higher the accretion rate in the disc, the lower the temperature and density in the corona. This leads to a small Compton $y-$parameter ($y=4kT_{\rm c}\tau_{\rm es}/m_{\rm e}c^2$ for optical-thin corona, where $T_{\rm c}$ is the temperature of the corona and $\tau_{\rm es}$ is the electron scattering depth), and a steep X-ray spectrum in luminous AGNs. Even all the accretion energy is efficiently transported from the disc to the corona by the magnetic field and subsequently emitted in the X-ray band, approximately half of the isotropic X-ray photons return to the disc and are mostly reprocessed as soft photons to take part in the Compton cooling again in the corona. Thus, the Compton $y-$parameter is anyhow not sufficiently large for producing a flat X-ray spectrum.

Some observational evidences suggest that certain luminous AGNs have much flat X-ray emission, i.e., $\Gamma_{\rm 2-10\,keV}=1.89 \pm0.05$ for radio-quiet quasars and $\Gamma_{\rm 2-10\,keV}\sim1.7-1.9$ for Seyfert 1 galaxies \cite[e.g.,][]{2000MNRAS.316..234R,2016MNRAS.458.2454L,2021A&A...655A..60A}. If these hard X-ray photons are emitted through the inverse Compton scattering of thermal electrons, the Compton $y$-parameter should exceed the theoretically predicted value.

In order to produce a flat spectrum as observed in radio-quiet quasars or Seyfert 1 galaxies, it is necessary to reduce the coronal illumination on the disc, resulting in a decrease in the energy of soft photons. This can be realized by assuming a truncated disc, the cold clumps, or an outflowing corona \citep{2000ApJ...542..703Z,2003ApJ...594L..99Y,2014ApJ...783..106L}. Nevertheless, the truncated disc is hard to produce a soft-state spectrum, and the cold clumps is to be further studied for the dynamical existence. Even though the radiation pressure-driven outflows could be optically thick with relatively low temperature, sustaining such an optically thick ``corona'' for steady X-ray emission is still questionable. On the other side, \cite{2020MNRAS.495.1158C} have argued that the reflection component flattens the X-ray spectrum with $\Delta\Gamma_{\rm 2-10\,keV}\sim 0.1-0.2$. The large reflection component is suggested to present in the Seyfert population \citep{2017ApJS..233...17R,2021A&A...655A..60A}. However, there is poor evidence for this component in quasars. Additionally, other accretion flows, such as those where coronal gas condenses into discs, are also suggested to produce relatively flat X-ray spectra \citep{2015ApJ...806..223L}. However, the current version of this model is only limited for the accretion rates below 0.1 times of the Eddington accretion rate.

  The electrons in the disc-corona system are commonly assumed to be thermal. However, we notice that some works have suggested that thermal and non-thermal electrons co-exist in the corona. The thermal and non-thermal electrons have a hybrid energy distribution. The investigations on the X-ray emission of hybrid electrons began in 1980s \citep{1983MNRAS.205..593G,1984ApJ...287..112K,1985ApJ...294L..79Z,1987MNRAS.227..403S,1987ApJ...319..643L}. In general, these works proposed that the $\gamma-\gamma$ process generates electron-positron pairs. This process takes effects on the optical depth of the corona and makes the result in changing X-ray emission. The inverse Compton scattering of coronal hybrid electrons was applied to the X-ray spectrum extending to $\sim800$ keV in the soft state of Cyg X-1, \citep[e.g.,][]{1999MNRAS.309..496G}. Without making any assumption about the shape of the particle distribution, \cite{2008A&A...491..617B} successfully solved the time-dependent kinetic equations for homogeneous, isotropic distributions of photons, electrons, and positrons. The particle heating and the radiation processes were fully considered. When the steady state is reached, non-thermal electrons are formed in the high-energy band. Thus, the emission properties of compact high energy sources can be well reproduced.
\cite{2009ApJ...690L..97P} and \cite{2009ApJ...698..293V} solved the coupled integro-differential kinetic equations for photons and electrons/positrons without any limitations on the photon and lepton energies. They reproduced the X-ray emission of Cyg X-1 in the soft state and suggested that the corona is magnetically dominated. In particular, magnetic reconnection in the context of accreting black-hole systems was considered by \cite{2017ApJ...850..141B} and \cite{2021MNRAS.507.5625S,2023MNRAS.518.1301S}. Recently, \cite{2022RAA....22c5002Z} suggested that the non-thermal electrons are accelerated by the first-order Fermi mechanism which is related to the magnetic-reconnection process. Therefore, in the corona, due to the continuous magnetic reconnection, some electrons can be accelerated to be non-thermal \cite[see][for review]{2015SSRv..191..545K}. The thermal and non-thermal electrons are expected to be co-existing presented by a hybrid energy distribution in the corona. Consequently, they can modify the radiation spectrum of the thermal electrons. The inverse Compton scattering of these non-thermal electrons in the hot corona can be used to explain the observed hard X-ray spectrum.

Although the radiation process of non-thermal electrons for X-ray emission has been well investigated, the coupling of the disc and the corona containing hybrid electrons needs to be further studied. We perform the magnetic-reconnection-heated corona model in this paper. The model has been well established \citep{2002ApJ...572L.173L,2003ApJ...587..571L,2016ApJ...833...35L,2020MNRAS.495.1158C}. In this work, the contribution of the radiation from the non-thermal electrons is included in the model, enabling us to successfully reproduce the flat X-ray spectrum ($\Gamma_{\rm 2-10\,keV}<2.1$).
In light of the coexistence between thermal electrons and non-thermal electrons in the magnetic reconnection process, we suppose that a fraction of the magnetic energy ($f_{\rm th}$) in the corona is allocated to thermal electrons, while the rest is allocated to non-thermal electrons. We show the dependence of the structure and spectrum of the disc-corona system on the energy allocated to the thermal electrons. We also investigate the effects of the non-thermal electrons and the magnetic field on the spectrum. Through comparison to the data of five individual luminous AGNs with $\Gamma_{\rm 2-10\,keV}< 2.1$ by our modeling results, we suggest that the Comptonization of hybrid electrons can well reproduce the flat X-ray spectra if a large fraction of magnetic energy ($>40\%$) is assigned to the non-thermal electrons.

This paper is organized as follows. In section 2, we introduce the model of disc and corona coupled by the magnetic field, which includes hybrid electrons in the corona. In section 3, we present the computational results on the structure and the spectrum of the disc-corona system in our model. In section 4, we apply the model for some individual objects in the hard X-ray band. The discussion and the conclusion are presented in Sections 5 and 6, respectively.

\section{The model} \label{sec:model}
 In this paper we adopt the magnetic-reconnection-heated corona model proposed by \cite{2002ApJ...572L.173L} and developed by \cite{2003ApJ...587..571L,2016ApJ...833...35L} and \citet{2020MNRAS.495.1158C}. In the model, the thin disc extending into the innermost stable orbit is sandwiched by the plane-parallel corona. The magnetic fields are generated by the dynamo mechanism in the accretion disc. Because of the magnetic buoyancy, the magnetic flux loops emerge into the corona at the Alfv\'{e}n speed $V_{\rm A}$ and reconnect with the opposite-direction loops. During the reconnection process, the magnetic energy is released as heating to the corona. The heating flux $Q_{\rm cor}^{+}$ in the corona can be expressed by the magnetic field $B$ and the Alfv\'{e}n speed $\rm V_A$ as
 \begin{eqnarray}\label{e:qcor}
Q_{\rm cor}^{+}={B^2\over 4\pi}V_{\rm A},
\end{eqnarray}
where the strength of the magnetic field $B$ is characterized by the magnetic coefficient $\beta_{\rm 0}$ as
 \begin{eqnarray}\label{eq:b}
 \beta_{\rm 0} = (P_{\rm g,d}+P_{\rm r,d})/(B^2/8\pi).
 \end{eqnarray}
 In the equation above, $P_{\rm g,d}=\frac{\rho_{\rm d}kT_{\rm d}}{\mu m_{\rm H}}$ and $P_{\rm r,d}=\frac{aT_{\rm d}^4}{3}$ are the gas-pressure and the radiation-pressure in the disc respectively, where $T_{\rm d}$ and $\rho_{\rm d}$ are the temperature and the density in the disc middle plane respectively, $k = 1.38\times10^{-16}\,\rm erg~K^{-1}$ is the Boltzmann constant, $m_{\rm H} = 1.67\times10^{-24}\,\rm g$ is the mass of the hydrogen atom, $\mu=0.5$ is the molecular weight for an
assumed chemical abundance of pure hydrogen, and $a =7.56\times10^{-15}\,\rm erg~K^{-4}~cm^{-3}$ is the radiation constant.

In the new scenario, non-thermal electrons are supposed to coexist with the thermal electrons in the corona, following a power-law distribution in the energy range between $\gamma_{\rm 1}$ and $\gamma_{\rm 2}$ as $n_{\rm e,pl}(\gamma)=C_{\rm 1}N_{\rm e,pl}\gamma^{-p}$, where $N_{\rm e, pl}$ is the  number density of the non-thermal electrons and $C_{\rm 1}\,=\,\frac{1-p}{\gamma_{\rm 2}^{1-p}-\gamma_{\rm 1}^{1-p}}$.

 The Alfv\'{e}n speed in Equation (\ref{e:qcor}) is a function of magnetic field and coronal density, $V_{\rm A}=B/\sqrt{4\pi\mu m_{\rm H}(N_{\rm e,th}+N_{\rm e,pl})}\approx B/\sqrt{4\pi\mu m_{\rm H}N_{\rm e,th}}$, where the number density of thermal electrons, $N_{\rm e,th}$, is usually much larger than that of non-thermal electrons, $N_{\rm e, pl}$,  as shown in section 5.2.

The ratio of the heating flux $Q_{\rm cor}^{+}$ to the total gravitational energy
%,denoted as $f$,
is defined as
\begin{eqnarray}\label{e:f}
f\equiv{Q_{\rm cor}^{+}\over Q_{\rm grav}}={{B^2\over 4\pi}V_{\rm A}\over {3GM_{\rm BH}\dot M \over 8\pi R^3}\left[{1-\left({3R_{\rm S}\over R}\right)^{1/2}}\right]},
\end{eqnarray}
where $R_{\rm s }\equiv 2GM_{\rm BH}/c^2$ is the Schwarzschild radius, $Q_{\rm grav}={3GM_{\rm BH}\dot M \over 8\pi R^3}\left[{1-\left({3R_{\rm S}\over R}\right)^{1/2}}\right]$ is the total accretion energy flux liberated in the accretion disc,
$R$ is the radius in the disc, $\dot M$ is the accretion rate, $\rm{M_{BH}}$ is the black hole mass, $c$ is the light speed,
%is $c = 3.00\times 10^{10}$\,\rm cm~s$^{-1}$},
and $G$ is the gravitational constant.
%  = 6.67\times10^{-8}\,\rm dyn~cm^2~g^{-2}$}.

  Since a fraction $f$ of the accretion energy is carried into the corona, the equation for the energy conservation in the thin disc is modified as
\begin{eqnarray}\label{eq:energy-d}
{3GM_{\rm BH}\dot M (1-f)\over 8\pi R^3}\left[{1-\left({3R_{\rm S}\over R}\right)^{1/2}}\right]
={4\sigma T_{\rm d}^4\over 3(\kappa_{\rm es}+\kappa_{\rm ff})H_{\rm d}},
\end{eqnarray}
where the scattering opacity is $\kappa_{\rm es}=0.4\,\rm ~{cm^{2}~g^{-1}}$, the free-free opacity is $\kappa_{\rm ff}=6.4\times10^{22}\rho_{\rm d}T_{\rm d}^{-7/2}\,\rm ~{cm^{2}~g^{-1}}$, and the Stefan-Boltzmann constant is $\sigma=5.67\times10^{-5}\,\rm erg~K^{-4}~cm^{-2}~s^{-1}$. The disc thickness is $H_{\rm d}=\sqrt{P_{\rm t, d}/\rho_{\rm d}}/{\Omega}$, where the total pressure is $P_{\rm t,d}=(1+1/\beta_0)(P_{\rm g,d}+P_{\rm r,d})$, and the angular velocity is $\Omega=\sqrt{GM_{\rm BH}/R^3}$.

 The corresponding angular momentum equation in the thin disc is
\begin{eqnarray}\label{eq:momentum-d}
 {\dot{M}\Omega\left[1-\left({\frac {3R_{\rm
s}}{R}}\right)^{1/2}\right](1-f)=4\pi H_{\rm d}\tau_{\rm r\varphi}},
\end{eqnarray}
where $\tau_{\rm r\varphi}=\alpha P_{\rm t,d}$ is the viscosity stress, and $\alpha$ is the viscosity parameter. From Equation (\ref{eq:energy-d}) and Equation (\ref{eq:momentum-d}), we can find that the temperature and the density in the disc are tightly related with the input parameters of $M_{\rm BH}$, $\dot M $, $\alpha$, and $\beta_{\rm 0}$. Moreover, they are also affected by the coronal density through the definition of the energy fraction $f$ in Equation (\ref{e:f}).

 The magnetic energy transferred from the disc is liberated in the corona via the process of the magnetic reconnection. If the density of the corona is not sufficiently high to radiate the gained energy, strong heat conduction to the chromospheric layer between the disc and the corona makes the gas evaporation into the corona. Once the density is reached a certain value, an equilibrium is established between magnetic heating and Compton cooling. Meanwhile, the evaporating gas steadily supplies for coronal accretion. Therefore, the corona and the disc are coupled in terms of energy and matter. Due to the coexistence of thermal and non-thermal electrons in the corona, a fraction ($f_{\rm th}$) of magnetic energy is suggested to heat the thermal electrons while the rest portion ($1-f_{\rm th}$) is used to accelerate the non-thermal electrons. The corona with certain density will be cooled by the inverse Compton scattering of the hybrid-distributed electrons. Finally, a steady corona with hybrid electrons is formed when the equilibrium is reached between the magnetic-reconnection heating and the Compton cooling. These coupling processes between the disc and the corona can be described using the following equations.

The energy balance equations for thermal and non-thermal electrons are
 \begin{eqnarray}\label{eq:corona1}
f_{\rm th} Q_{\rm cor}^+ \approx \frac{4kT_{\rm c}}{m_{\rm e}c^2} \lambda_{\rm \tau} N_{\rm e,th}\sigma_{\rm T}\ell_{\rm c}cU_{\rm rad},
 \end{eqnarray}
and
\begin{eqnarray}\label{eq:corona2}
 (1-f_{\rm th})Q_{\rm cor}^+ & \approx & \int_{\gamma_{\rm 1}}^{\gamma_{\rm 2}}\frac{4}{3}\sigma_{\rm T}c\gamma^2\beta^2 U_{\rm rad}\lambda_{\rm \tau}\ell_{\rm c}n_{\rm e,pl}(\gamma)d\gamma \nonumber \\
 & =& 4C_{\rm 1}G(\gamma_{\rm 1},\gamma_{\rm 2},p)\lambda_{\rm \tau}N_{\rm e,pl}\sigma_{\rm T}\ell_{\rm c} cU_{\rm rad},
\end{eqnarray}
 where $G(\gamma_{\rm 1}, \gamma_{\rm 2},p) = \frac{1}{3}\int_{\rm \gamma_{\rm 1}}^{
  \gamma_{\rm 2}}\gamma^{2-p}\beta^2d\gamma$.

 The thermal conduction-induced evaporation as
\begin{eqnarray}\label{eq:evap}
{k_0T_{\rm c} ^{7\over 2}\over \ell_{\rm c}}\approx
{\gamma_{\rm 0}\over \gamma_{\rm 0}-1} N_{\rm e,th} k T_{\rm c}
\left(\frac{kT_{\rm c}} {\mu m_{\rm H}}\right)^{1/2},
\end{eqnarray}
where $T_{\rm c}$ is the temperature of the thermal electrons. The constants in the above equations are: the ratio of specific heats $\gamma_{\rm 0}=5/3$ and the thermal conduction coefficient $k_0=10^{-6}\,\rm {erg~cm^{-1}~s^{-1}~K^{-7/2}}$. The length of the magnetic loops in the corona $\ell_{\rm c}$ is approximate to $R$ for the optical-thin corona. As the isotropic incident photons are up-scattered in the plane-parallel corona, in Equations (\ref{eq:corona1}) and (\ref{eq:corona2}), we introduce a coefficient $\lambda_{\rm \tau}$, which is generally larger than 1. The energy density of soft photons, $U_{\rm rad}$, is contributed by the intrinsic disc emission released by the viscous heating and the reprocessing of the coronal irradiation. We present it in the following.

The inverse Compton scattering of the isotropically distributed electrons implies that about half of the Comptonization photons upward-escape from the corona, while the rest photons are backward to the disc. Within only a small fraction undergoing reflection (albedo $a=0.2$), these backward photons are reprocessed as seed photons for Compton cooling in the corona. Therefore, the soft photon flux for coronal inverse Compton scattering is $F_{\rm s}=(1-f)Q_{\rm grav}+\frac{1}{2}(1-a)F_{\rm c}$, where $F_{\rm c}$ is the total Compton emission flux from both the thermal electrons ($F_{\rm th}$) and the non-thermal electrons ($F_{\rm  pl}$). The thermal flux $F_{\rm th}$ can be expressed as the sum of the net flux gained during scattering and the original flux brought by the scattered soft photons, i.e., $F_{\rm  th}=f_{\rm th}fQ_{\rm grav}+F_{\rm s}(1-e^{-\tau_{c}})$, where $\tau_{c}\equiv\lambda_{\rm \tau}N_{\rm e,th}\sigma_{\rm T}\ell_{\rm c}$ is the effective optical depth and $(1-e^{-\tau_{c}})$ is the scattering probability of a soft photon. The flux $F_{\rm  pl}$ is corresponding to the magnetic energy allocated to the non-thermal electrons, i.e., $F_{\rm pl}=(1-f_{\rm th})fQ_{\rm grav}$. Therefore, the flux and the energy density for the soft photons can be expressed as
  \begin{eqnarray} \label{eq:soft-flux}
    F_{\rm s}=\frac{1-f+\frac{1}{2}(1-a)f}{1-\frac{1}{2}(1-a)(1-e^{-\tau_{\rm c}})}Q_{\rm grav} { \   \   \   } \rm{and} { \   \   \   } U_{\rm rad}=2F_{\rm s}/c,
  \end{eqnarray}
  respectively.

  For given values of the black hole mass $M_{\rm BH}$, the accretion rate $\dot{M}$, the viscosity parameter $\alpha$, the magnetic coefficient $\beta_{\rm 0}$, the fraction of magnetic energy to thermal electrons $f_{\rm th}$, and the parameters for the non-thermal electrons including $\gamma_{\rm 1}$, $\gamma_{\rm 2}$ and $p$, we can numerically solve the Equations (\ref{eq:energy-d}), (\ref{eq:momentum-d}), (\ref{eq:corona1}), (\ref{eq:corona2}), and
  (\ref{eq:evap}) by combing Equations (\ref{e:qcor}), (\ref{eq:b}), (\ref{e:f}), and (\ref{eq:soft-flux}) at the grid point of radius R, with an initial value of $\lambda_{\rm \tau}=1.0$. The temperature $T_{\rm d}$ and the density $\rho_{\rm d}$ in the disc, the temperature $T_{\rm c}$ and the number density $N_{\rm e,th}$ of thermal electrons, and the number density $N_{\rm e,pl}$ of non-thermal electrons in the corona are subsequently derived. The values of $f$ and $U_{\rm rad}$ can be also obtained according to their definitions in Equation (\ref{e:f}) and  Equation (\ref{eq:soft-flux}). Utilizing the structural parameters $T_{\rm c}$, $N_{\rm e,th}$ and $U_{\rm rad}$ of the disc-corona system, we can calculate the X-ray spectrum emitted by thermal electrons through Monte Carlo simulation. We need to check the self-consistency from the Monte Carlo simulation result, i.e., whether the flux of escaped photons in the Monte Carlo simulation is equal to the combined flux from upward Compton emission and directly escaped (non-scattered) soft photons, i.e., ${F_{\rm th}/2}+F_{\rm s}e^{-\tau_{\rm c}}$. In order to fulfill this condition, we fine-tune the parameter $\lambda_{\rm \tau}$ and repeat the calculation for the disc-corona structure and the Monte Carlo simulation. Subsequently, we update the value of $R$ to the adjacent grid point to computer the structure and spectrum of the disc-corona system. Ultimately, we determine the global structural parameters of the disc-corona system and the overall spectrum contributed by thermal electrons.

In order to calculate the radiation spectrum of non-thermal electrons, we use the modified Compton spectrum formalism in the Thomson limit proposed by \citet{2013ApJ...773...23Z}. The presentation is
\begin{eqnarray}
\frac{dN_{ \gamma,\epsilon} (R)}{dtd\epsilon_1}=\frac{\pi r_0^2 c}{2\gamma^4 \beta^4}\frac{n_{\rm ph}(\epsilon,R)d\epsilon}{\epsilon}
\varsigma(\frac{\epsilon_{\rm 1}}{\epsilon},\gamma),
\end{eqnarray}
where the classical electron radius is $r_{\rm 0}= 2.82\times10^{-13}$\,cm and $\varsigma(\frac{\epsilon_{\rm 1}}{\epsilon},\gamma)=2\frac{\epsilon_{\rm 1}}{\epsilon} \rm{ln}\frac{\epsilon_1}{\epsilon(1+\beta)^2 \gamma^2 }+\frac{2\beta}{1+\beta}
\frac{\epsilon_1}{\epsilon}+(1+\beta)(1+\beta^2)\gamma^2-
\left(\frac{\epsilon_1}{\epsilon}\right)^2\frac{1}{(1+\beta)\gamma^2}$.
The parameters $\epsilon$ and $\epsilon_{\rm 1}$ represent the energy of the initial photon and the energy of the photon after scattering, respectively. The differential density of the seed photons at the radius $R$ is
  \begin{eqnarray}
 n_{\rm ph}(\epsilon,R)d\epsilon =\frac{4\pi\epsilon^2}{h^3c^3(e^{\frac{
 \epsilon}{kT_{\rm eff}(R)}}-1)}d\epsilon,
\end{eqnarray}
where $T_{\rm eff}(R)= cU_{\rm rad}/(2\sigma)$ is the effective temperature at the radius $R$ of the thin disc.
Finally, the total radiation rate emitted by the non-thermal electrons per volume is
 \begin{eqnarray}\label{e:nth-radiation}
\frac{dN_{\rm tot}(R)}{dtd\epsilon_1} &=&\int_{\epsilon_{0}}^ {\epsilon_{1}}\int_{\rm \gamma_{\rm 1}}^{\gamma_{\rm 2}} n_{\rm e,pl}(\gamma)\frac{dN_{\rm \gamma,\epsilon}(R)}{dtd\epsilon_1}
d\epsilon d\gamma \nonumber\\
&=&\frac{\pi r_{\rm 0}^{2}c}{2}\int_{\epsilon_{0}}^ {\epsilon_{1}}
\int_{\rm \gamma_{\rm 1}}^{\gamma_{\rm 2}}\frac{n_{\rm ph}(\epsilon,R)}{\epsilon}\times\nonumber \\
& & N_{\rm e,pl}C_{\rm 1} \frac{\varsigma(\frac{\epsilon_{\rm 1}}{\epsilon},\gamma)}{\gamma^{p+4}\beta^4}d\epsilon d\gamma,
\end{eqnarray}
 where the lower-energy limit is $\epsilon_{\rm 0}=\epsilon_{\rm 1}(1-\beta_{\rm 2})/(1+\beta_{\rm 2})$ with $\beta_{\rm 2}=\sqrt{1-1/\gamma_{\rm 2}^2}$.

We can calculate the total luminosity of electrons in the corona as
\begin{eqnarray}
 L(\epsilon_{\rm 1}) & = & \epsilon_{\rm 1}^2\int_{R_{\rm in}}^{R_{\rm out}} 4\pi R\ell_{\rm c}\frac{dN_{\rm tot}(R)}{dtd\epsilon_{\rm 1}}dR  \nonumber\\
 & \approx &  2(\pi\epsilon_{\rm 1}r_{\rm 0} )^2 \int_{R_{\rm in}}^{R_{\rm out}}\int_{\epsilon_{0}}^ {\epsilon_{1}}
\int_{\rm \gamma_{\rm 1}}^{\gamma_{\rm 2}}\frac{n_{\rm ph}(\epsilon,R)}{\epsilon}\times\nonumber \\
 &  &  \frac{N_{\rm e,pl}C_{\rm 1}\varsigma(\frac{\epsilon_{\rm 1}}{\epsilon},\gamma)}{\gamma^{p+4}\beta^4} R^2 dR d\epsilon d\gamma,
  \end{eqnarray}
where $R_{\rm in}$ and $R_{\rm out}$ are the inner boundary and outer boundary of the accretion disc-corona system, respectively. As approximately half of the Compton photons are backward to the disc, the radiation luminosity from the coronal upper layer is also half of the total luminosity.

\section{Numerical Results}\label{sec:results}
  In our previous works \citep{2016ApJ...833...35L,2020MNRAS.495.1158C}, we have shown the influence of the accretion rate and black hole mass on both the structure and spectrum. In this paper, we focus on the effect of the Comptonization of non-thermal electrons on the X-ray spectrum in luminous AGNs. Thus, we fix the black hole mass $M_{\rm BH}=10^8M_{\rm \odot}$ and the accretion rate $\dot{M}=0.1\dot{M}_{\rm Edd}$. The viscosity parameter $\alpha$ is 0.3. We change the energy fraction $f_{\rm th}$, the parameters ($\gamma_{\rm 1}$, $\gamma_{\rm 2}$, $p$) for the distribution of non-thermal electrons, and the magnetic coefficient $\beta_{\rm 0}$ in order to investigate their effects on the properties of the structure and the spectrum of the disc-corona system.

 \subsection{The effect of the energy fraction $f_{\rm th}$}\label{subs:fth}
\begin{figure}
\includegraphics[scale=0.4]{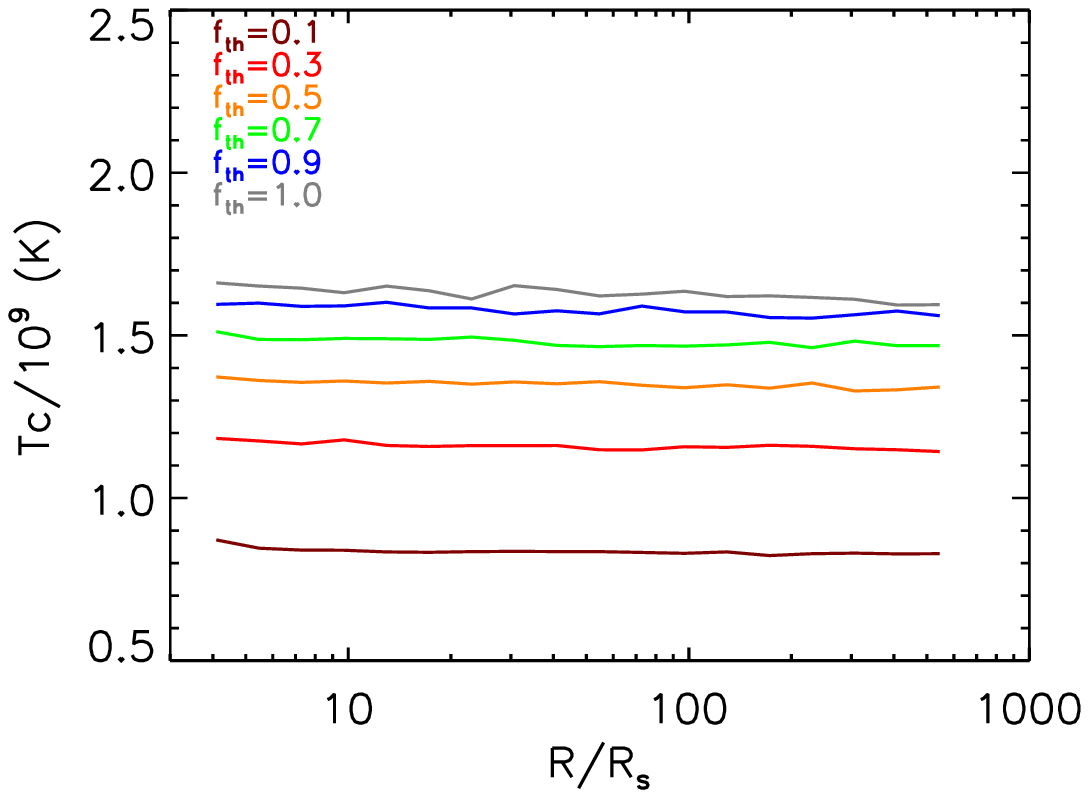}
\includegraphics[scale=0.4]{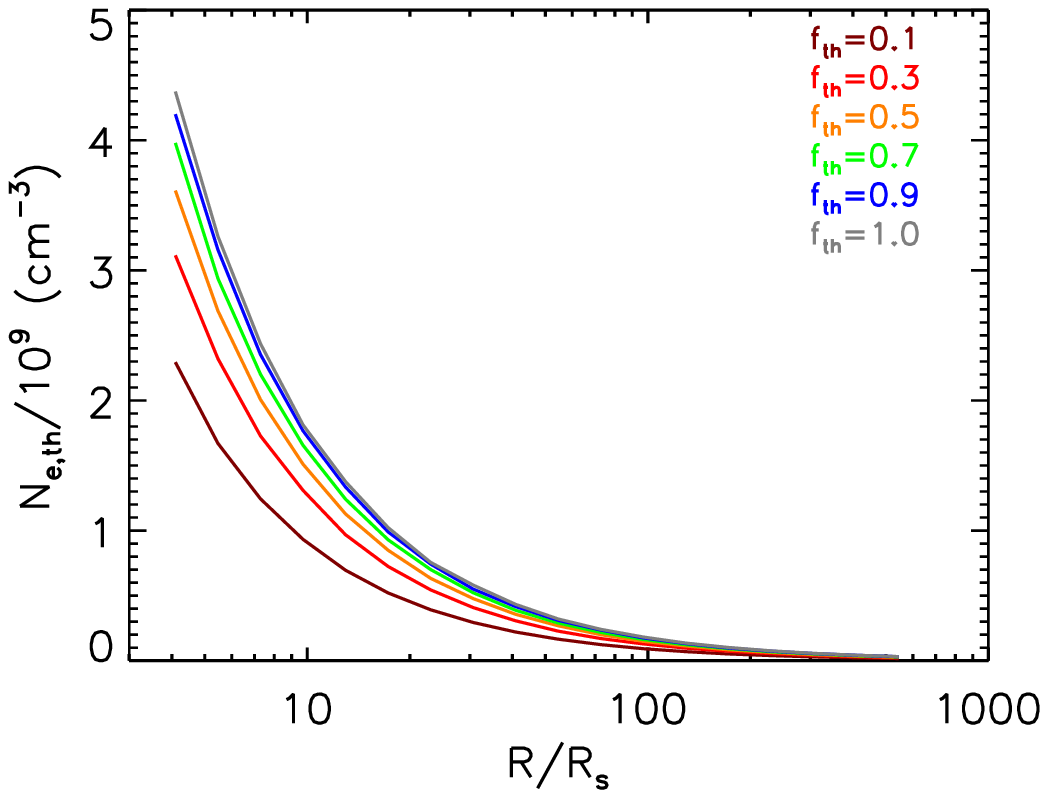}
\includegraphics[scale=0.4]{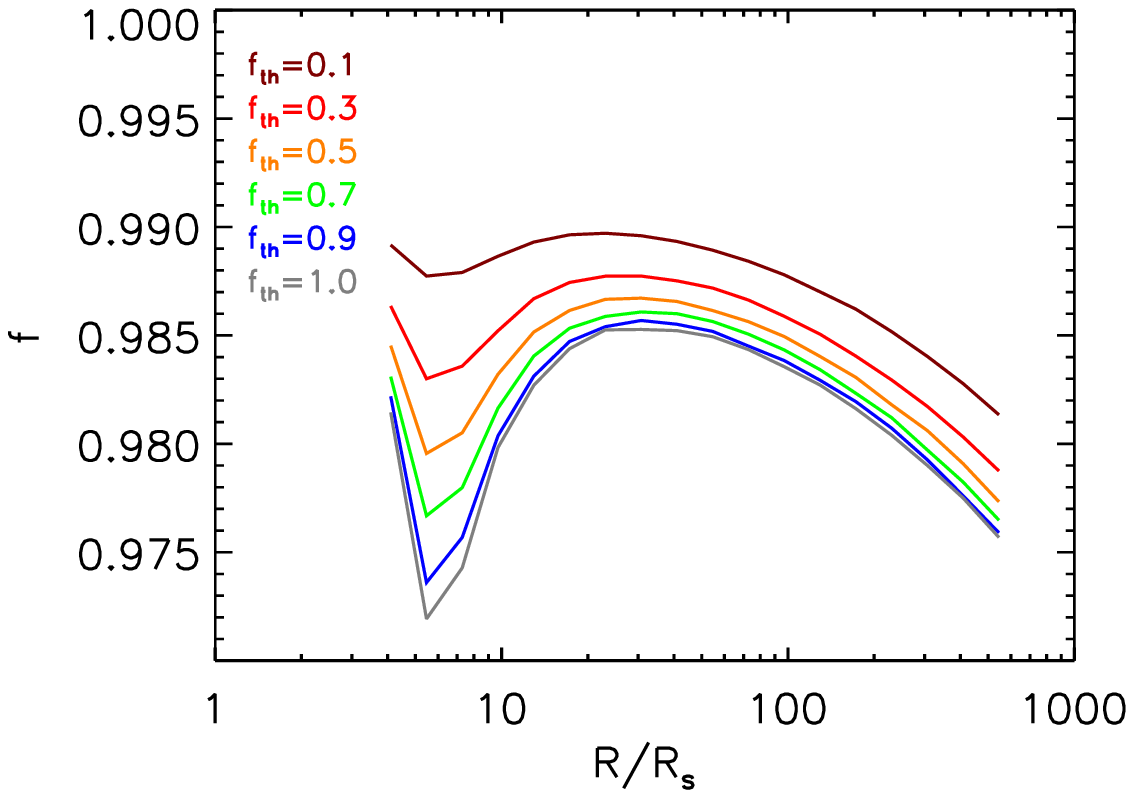}
\caption{The radial distribution of the temperature $T_{\rm c}$ (top panel), the density $N_{\rm e,th}$ of thermal electrons (middle panel), and the total energy fraction $f$ (bottom panel) with different values of $f_{\rm th}$. The grey lines in the figure show the cases where only thermal electrons are included in the corona.}\label{fig:fth-structure}
\end{figure}

  Firstly, we fix the magnetic field with $\beta_{\rm 0}=50$. In this case, approximately $97\%$ of the total gravitational energy is carried into the corona. We change the value of $f_{\rm th}$ to investigate its effects on the temperature, the density of thermal electrons, and the total energy fraction $f$. As shown in Figure \ref{fig:fth-structure}, the electrons are heated to a high temperature about $10^9$\,K.
  The temperature $T_{\rm c}$ and the density $N_{\rm e,th}$ of thermal electrons are decreased as the parameter $f_{\rm th}$ is decreased. Meanwhile, the total energy fraction $f$ is slightly increased when the parameter $f_{\rm th}$ is slightly decreased.

%%% figure 1 ============

%%% figure 2 ============
    \begin{figure}
     \includegraphics[scale=0.4]{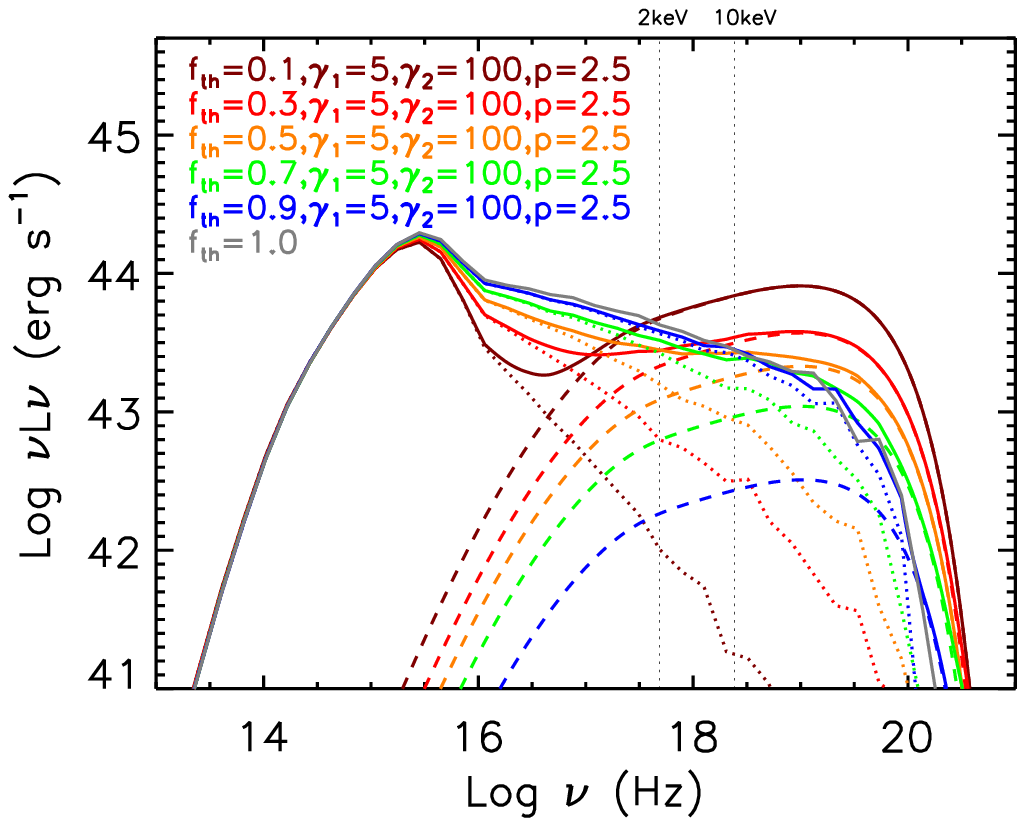}
     \includegraphics[scale=0.4]{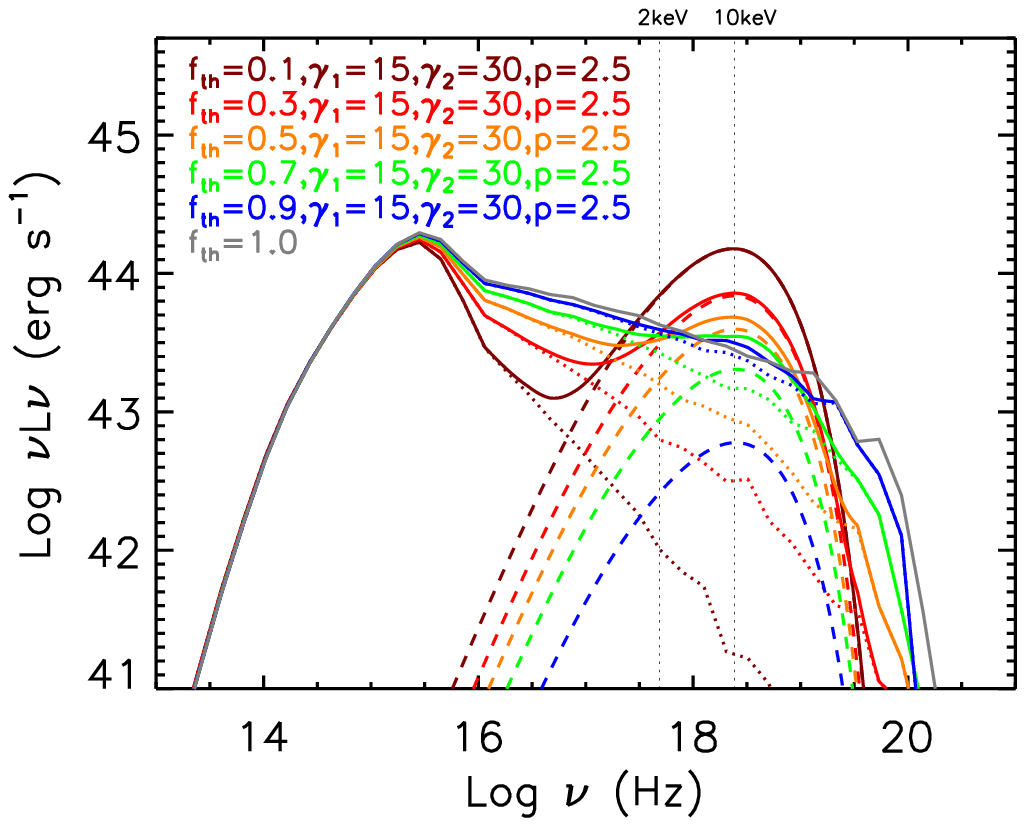}
   \caption{The spectra of the disc-corona system with different $f_{\rm th}$. The top panel: the non-thermal electron distribution is adopted with $\gamma_{\rm 1}=5$, $\gamma_{\rm 2}=100$, and $p=2.5$. The bottom panel: the non-thermal electrons with $\gamma_{\rm 1}=15$, $\gamma_{\rm 2}=30$, and $p=2.5$. The thick dotted lines represent the radiation spectra of thermal electrons, the dashed lines represent the radiation spectra of non-thermal electrons, and the solid lines represent the total radiation spectra of disc-corona system. The grey lines represent the radiation spectra of the disc-corona containing only thermal electrons. The two vertical dotted lines indicate the photon energies of 2\,keV and 10\,keV, respectively.}\label{fig:fth-spectrum}
\end{figure}

We show the spectra of the disc-corona with different $f_{\rm th}$ in Figure \ref{fig:fth-spectrum}. Two different energy distributions of the non-thermal electrons are adopted. The top panel is shown for the non-thermal electrons of $\gamma_{\rm 1}=5$, $\gamma_{\rm 2}=100$, and $p=2.5$. The bottom panel is shown for the non-thermal electrons of $\gamma_{\rm 1}=15$, $\gamma_{\rm 2}=30$, and $p=2.5$. The hard X-ray spectra in both panels show a trend towards flatter shape as more magnetic energy is allocated to non-thermal electrons. We note that the X-ray emission in the $2$--$10$\,keV energy band is dominated by the radiation from non-thermal electrons (indicated by the dashed lines in Figure \ref{fig:fth-spectrum}) when $f_{\rm th}$ is less than $0.5$.

\begin{figure}
 \begin{center}
 \includegraphics[scale=0.4]{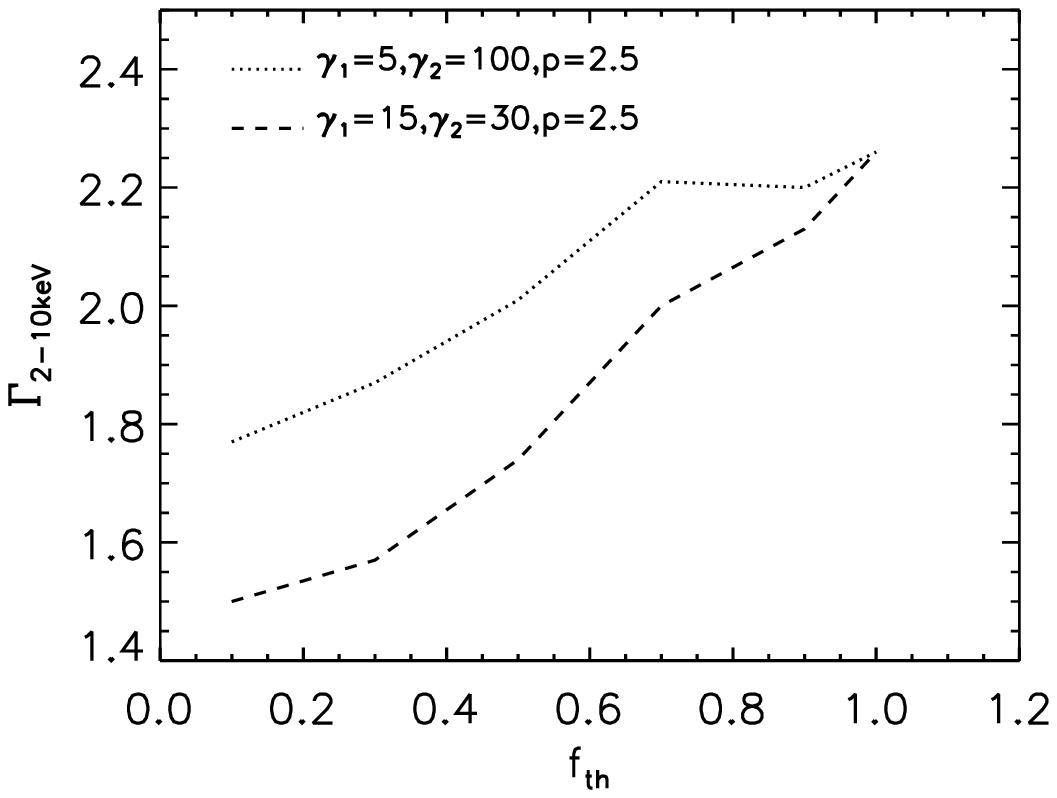}
 \caption{The photon index $\Gamma_{\rm 2-10\,keV}$ as a function of energy fraction $f_{\rm th}$. Two different distributions of non-thermal electrons are adopted: One is of $\gamma_{\rm 1}=5$, $\gamma_{\rm 2}=100$, and $p=2.5$ (dotted), the other is of $\gamma_{\rm 1}=15$, $\gamma_{\rm 2}=30$, and $p=2.5$ (dashed).}\label{fig:fth-gama}
 \end{center}
\end{figure}

The dependence of photon index $\Gamma_{\rm 2-10\,keV}$ on $f_{\rm th}$ is shown in Figure \ref{fig:fth-gama}. We find that $\Gamma_{\rm 2-10\,keV}$ is increased as $f_{\rm th}$ is increased. For the case of $\gamma_{\rm 1}=5$, $\gamma_{\rm 2}=100$, and $p=2.5$ (indicated by the dotted line), the disc-corona system with $f_{\rm th} < 0.6$ can reproduce the flat spectrum with $\Gamma_{\rm 2-10\,keV} < 2.1$. While, for the case of $\gamma_{\rm 1}=15$, $\gamma_{\rm 2}=30$, and $p=2.5$ (indicated by the dashed line), the disc-corona system with $f_{\rm th} < 0.9$ can reproduce the flat spectrum with $\Gamma_{\rm 2-10\,keV} < 2.1$. Furthermore, we can also find that the X-ray spectrum is significantly influenced by the energy distribution of non-thermal electrons, even for the same energy fraction $f_{\rm th}$.

%%% figure 3 ============

\subsection{The Effect of the Energy Range of Non-thermal electrons}\label{sec:g1}
  We fix $f_{\rm th}= 0.5$ and $p\,=\,2.5$, and we change the value of $\gamma_{\rm 1}$ to study the effect on the spectrum of the disc-corona system. Meanwhile, we consider two cases for comparison: one is for $\gamma_{\rm 2}=100$, while the other is for $\gamma_{\rm 2}=30$.

\begin{figure}
 \includegraphics[scale=0.4]{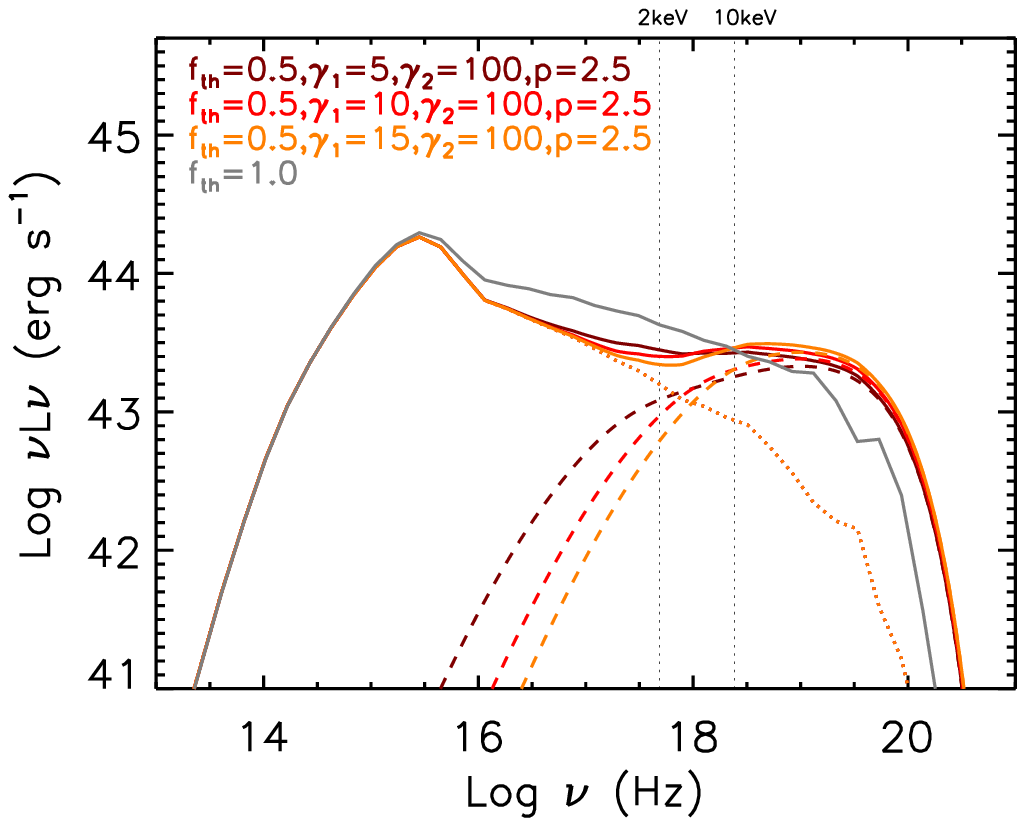}
 \includegraphics[scale=0.4]{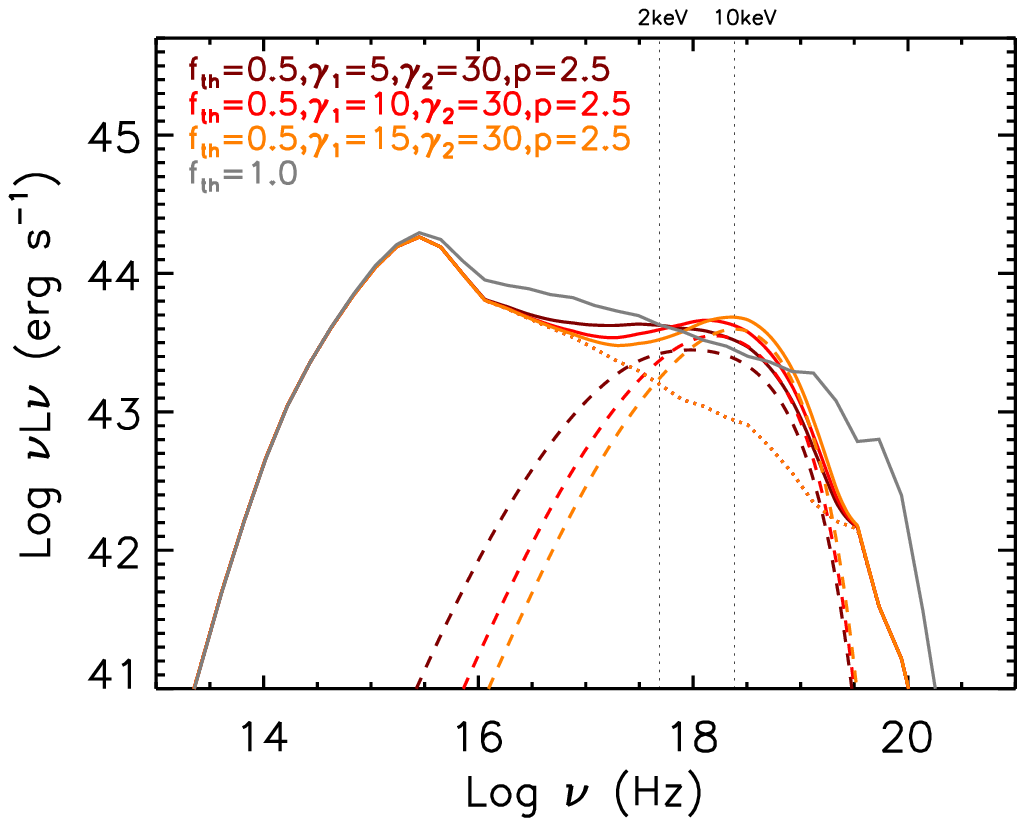}
 \caption{The spectra of the disc-corona system with the different $\gamma_{\rm 1}$ numbers (brown: $\gamma_{\rm 1}=5$; red: $\gamma_{\rm 1}=10$; orange: $\gamma_{\rm 1}=15$). $\gamma_{\rm 2}$ in the top and bottom panels are 100 and 30, respectively. The indications of the different line styles are the same as those in Figure \ref{fig:fth-spectrum}.}\label{fig:g1-spectrum}
\end{figure}

The spectra affected by different values of $\gamma_{\rm 1}$ are shown in Figure \ref{fig:g1-spectrum}. It is found that the spectra become flat when $\gamma_{\rm 1}$ is increased. For comparison, we consider two cases: one is for $\gamma_{\rm 2} = 100$ (top panel), the other is for $\gamma_{\rm 2}=30$ (bottom panel). For the case of $\gamma_{\rm 2} = 100$, the X-ray photon index is $\Gamma_{\rm 2-10\,keV}$=2.0 for $\gamma_{\rm 1}$=5, $\Gamma_{\rm 2-10\,keV}$=1.9 for $\gamma_{\rm 1}$= 10, and $\Gamma_{\rm 2-10\,keV}$= 1.8 for $\gamma_{\rm 1}$=15. For the case of $\gamma_{\rm 2}=30$, the X-ray photon index $\Gamma_{\rm 2-10\,keV}$ varies from 2.1 to 1.7 as the value of $\gamma_{\rm 1}$ is increased from 5 to 15.
%%% figure 4 ============

\begin{figure}
 \includegraphics[scale=0.4]{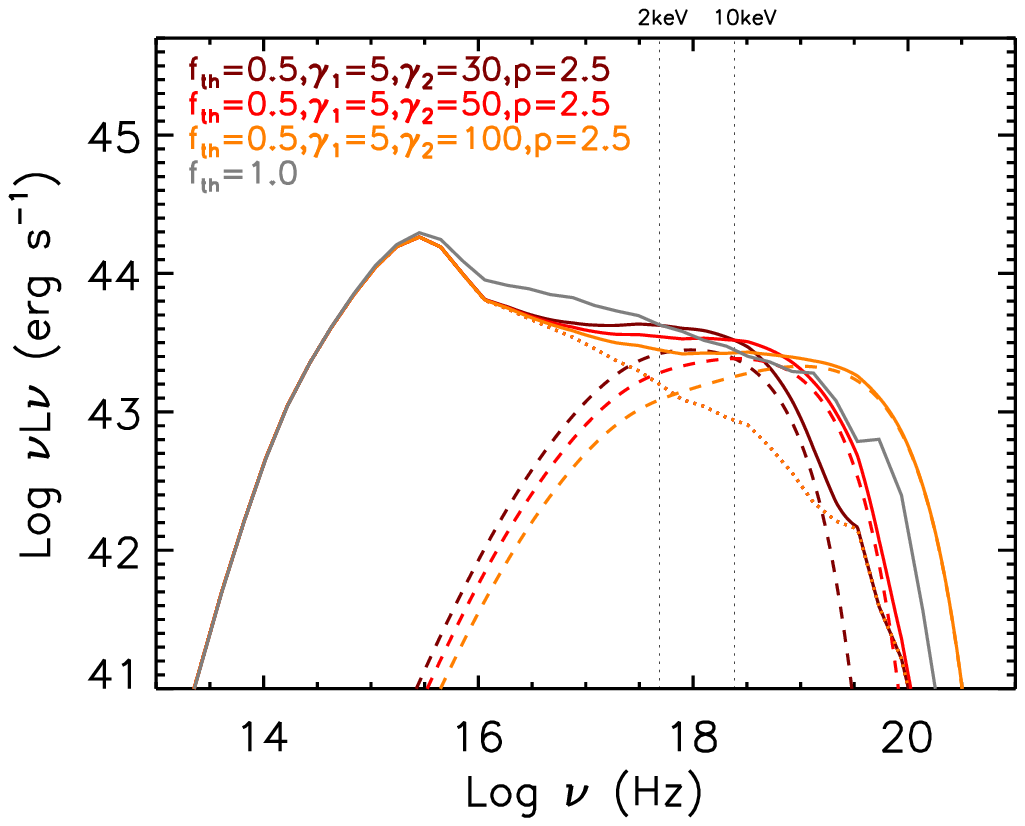}
 \includegraphics[scale=0.4]{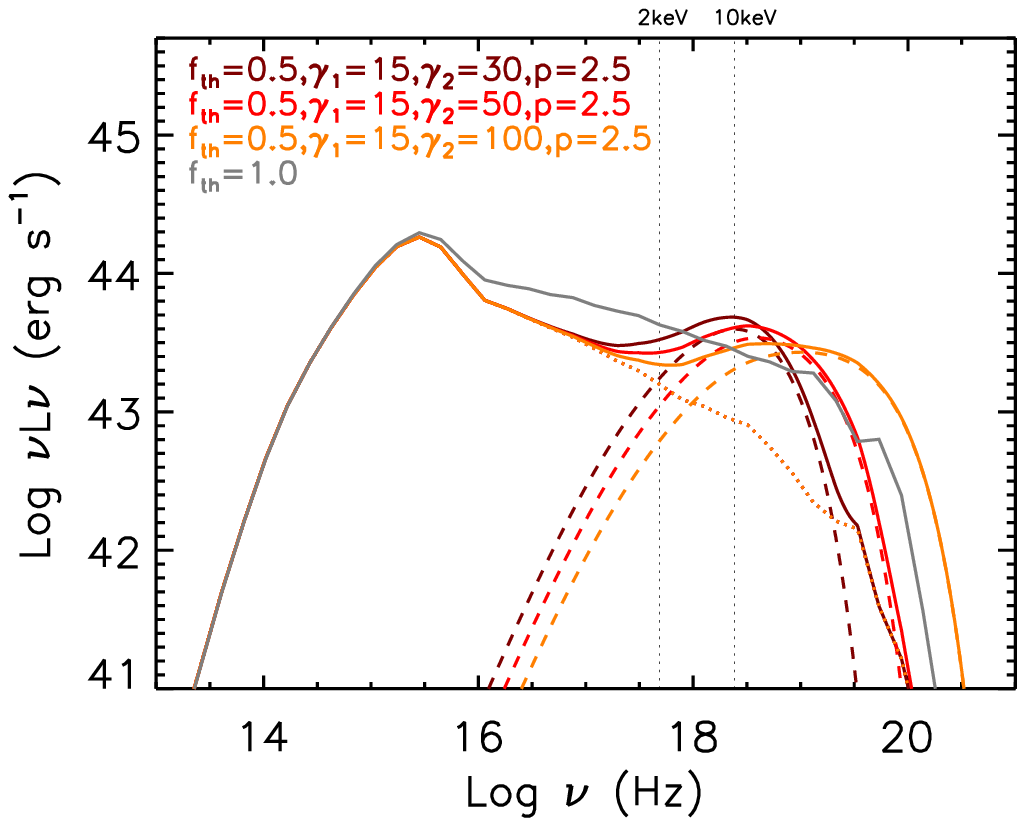}
 \caption{The radiation spectra of the disc-corona system with the different $\gamma_{\rm 2}$ numbers (brown: $\gamma_{\rm 2} =30$; red: $\gamma_{\rm 2}=50$; orange: $\gamma_{\rm 2}=100$). $\gamma_{\rm 1}$ in the top and the bottom panels are 5 and 15, respectively. The indications of the different line styles are the same as those in Figure \ref{fig:fth-spectrum}.}\label{fig:g2-spectrum}
\end{figure}

We also fix $f_{\rm th}=0.5$ and $p\,=\,2.5$, and change the value of $\gamma_{\rm 2}$ to investigate the effect on the spectrum of the disc-corona system. The spectra of the disc-corona system with different values of $\gamma_{\rm 2}$ are shown in Figure \ref{fig:g2-spectrum}. For comparison, we consider two cases: one is for $\gamma_{\rm 1}=5$, the other is for $\gamma_{\rm 1}=15$. For the case of $\gamma_{\rm 1}=5$, we obtain the similar X-ray photon indices for different values of $\gamma_{\rm 2}$. While, for the case of $\gamma_{\rm 1}=15$, $\Gamma_{\rm 2-10\,keV}$ is increased slightly from 1.7 to 1.8 when $\gamma_{\rm 2}$ is increased from 30 to 100.

A larger $\gamma_{\rm 2}$ number implies that soft photons can be up-scattered to higher energy-band. As more energy is emitted in higher energy-band for a larger $\gamma_{\rm 2}$ number, the X-ray luminosity %in $2$--$10$\,keV
is decreased. In particular, the non-thermal electrons in narrow energy range, such as $\gamma_{\rm 1}=15$ and $\gamma_{\rm 2} =30$ in the bottom panel of Figure \ref{fig:g2-spectrum}, will result in a ``bump'' emission in the X-ray band. The peak of the ``bump'' shifts towards a lower energy band as $\gamma_{\rm 2}$ is decreased. This implies that setting a narrow range of Lorentz factor and a much lower $\gamma_{\rm 2}$ number can generate the ``bump'' in the soft X-ray band below 2.0\,keV. It is consistent with the results found in \cite{2013ApJ...773...23Z}. In that paper, it was suggested that the soft X-ray excess originates from the Comptonization of the non-thermal electrons with a small number of $\gamma_{\rm 2}$.

 %%% figure 5 ============

\subsection{The Effect of the Electron Power-Law Index}\label{sec:p}
\begin{figure}
 \includegraphics[scale=0.4]{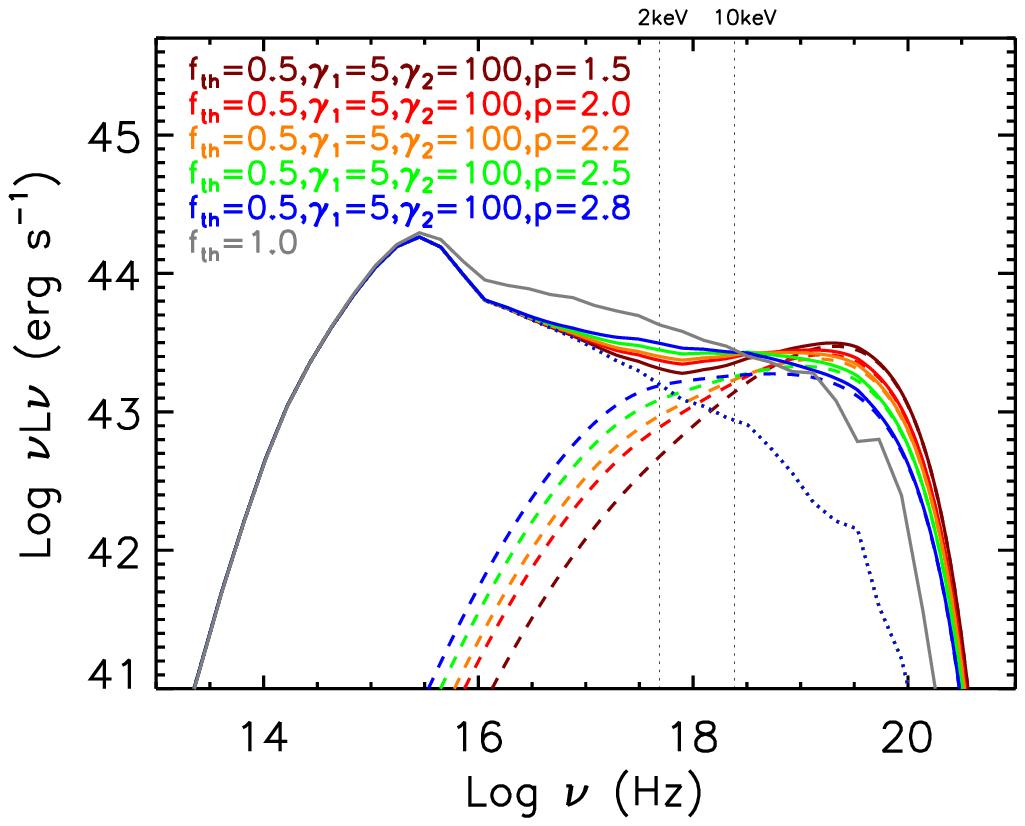}
   \includegraphics[scale=0.4]{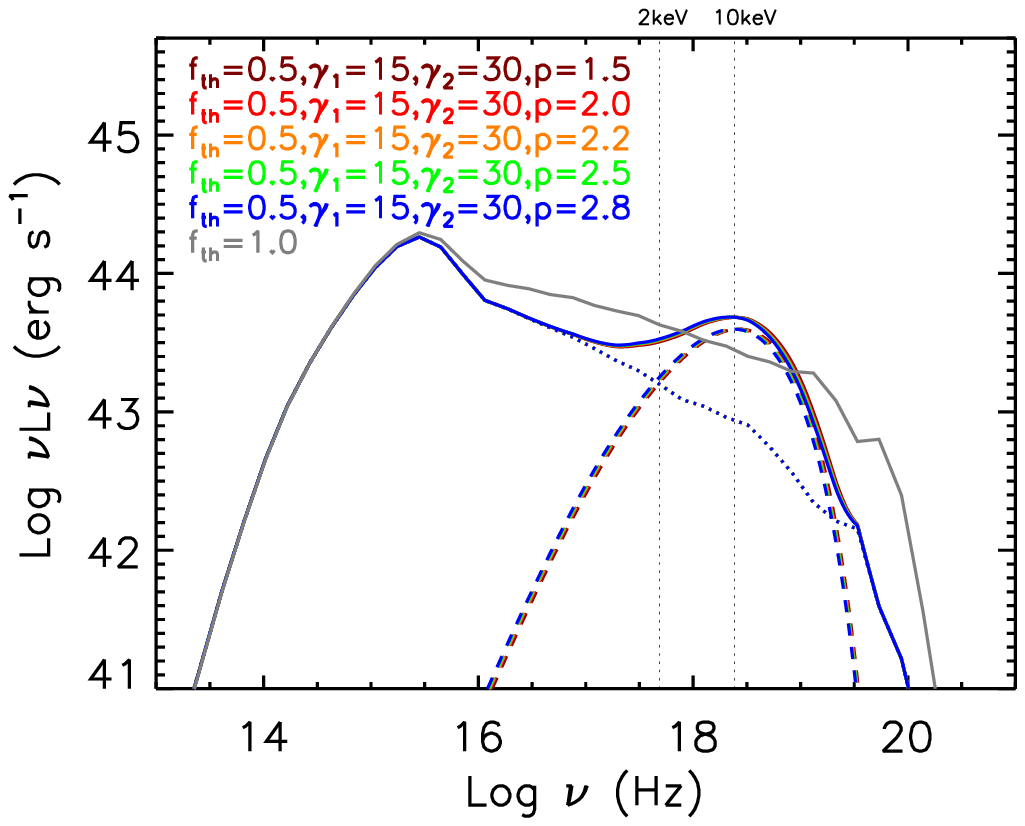}
 \caption{The radiation spectra of the disc-corona with the different $p$ numbers (brown: $p=1.5$; red: $p=2.0$; orange: $p=2.2$; green: $p=2.5$; blue: $p=2.8$). The indications of the different line styles are the same to those in Figure \ref{fig:fth-spectrum}.}\label{fig:p-spectrum}
\end{figure}

As shown in Figure \ref{fig:fth-gama}, because $\gamma_{\rm 2}$ is not significantly larger than $\gamma_{\rm 1}$, the calculated X-ray photon index is not simply approximated to $\frac{p+1}{2}$. In this subsection, we change the value of the electron power-law index $p$ from 1.5 to 2.8 in order to investigate its effect on the spectrum. The results are shown in Figure \ref{fig:p-spectrum}.

   For the case of [$\gamma_{\rm 1}$, $\gamma_{\rm 2}$]=[5, 100] shown in the top panel of Figure \ref{fig:p-spectrum}, the spectrum dramatically becomes flat as $p$ is decreased. While for the case of [$\gamma_{\rm 1}$, $\gamma_{\rm 2}$]=[15, 30] shown in the bottom panel, the spectrum has a ``bump'' signature and it is hardly affected by changing the value of $p$.

 %%% figure 6 ============

 \subsection{The Effect of the Magnetic Field}\label{sec:beta}
\begin{figure}
          \begin{center}
         \includegraphics[scale=0.4]{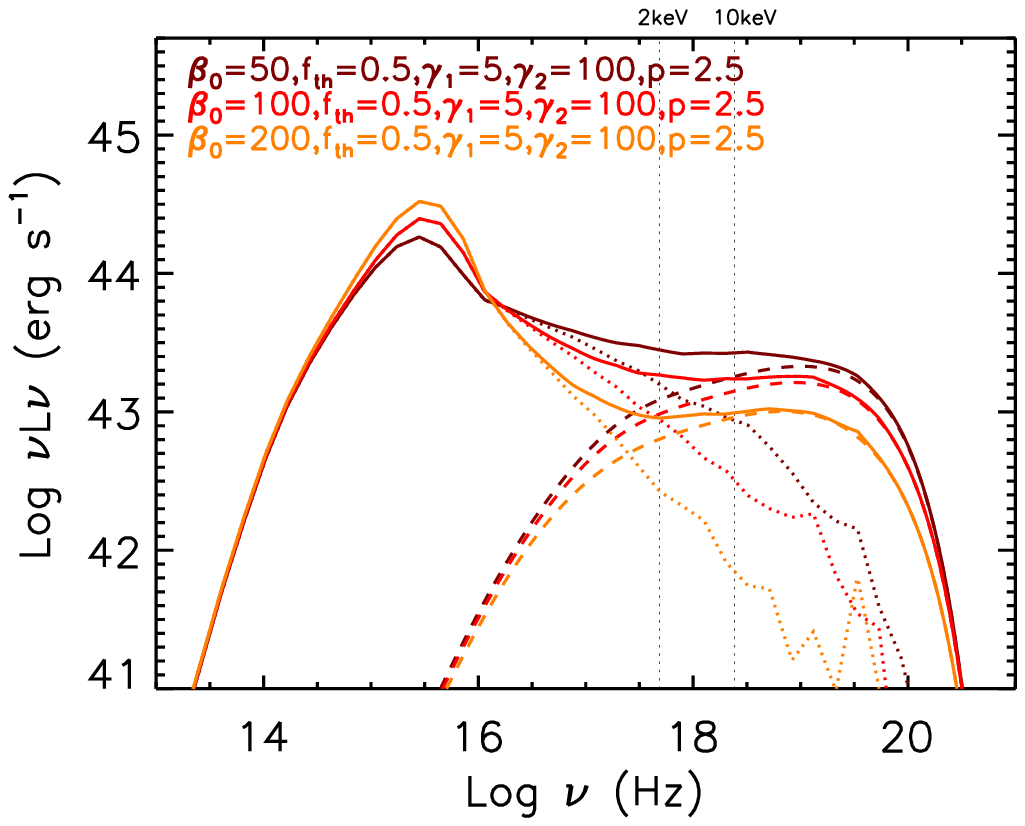}%
  \caption{The radiation spectra of the disc-corona with the different $\beta_{\rm 0}$ numbers (brown: $\beta_{\rm 0}=50$; red: $\beta_{\rm 0}=100$; orange: $\beta_{\rm 0}=200$).}\label{fig:beta-spectrum}
     \end{center}
\end{figure}

 Magnetic field plays a crucial role in our model. It is characterized by the magnetic coefficient $\beta_{\rm 0}$. A larger $\beta_{\rm 0}$ value represents a weaker magnetic field and less gravitational energy carried into the corona \citep{2016ApJ...833...35L,2020MNRAS.495.1158C}. We fix $f_{\rm th}=0.5$, $\gamma_{\rm 1}=5$, $\gamma_{\rm 2}=100$, and $p=2.5$. Then we change the value of $\beta_{\rm 0}$ to study the effect of the magnetic field on the radiation spectrum of the disc-corona system. As shown in Figure \ref{fig:beta-spectrum}, for the case of a weak magnetic field with $\beta_{\rm 0}=200$ (orange lines), a weak corona is built up and the X-ray emission is also weak. It is shown that the X-ray radiation contributed by the inverse Compton scattering of the hybrid electrons is increased as the magnetic field is increased. It is found that the X-ray emission is actually dominated by the Comptonization of non-thermal electrons (shown by the dashed lines in Figure \ref{fig:beta-spectrum}). Due to the identical energy distribution of the non-thermal electrons, we obtain a consistent X-ray photon index $\Gamma_{\rm 2-10\,keV}\sim 2.0$ across different magnetic coefficient values. Meanwhile, we note that the hard X-ray bolometric correction factor is increased as the magnetic field is decreased: $L_{\rm bol}/L_{\rm 2-10\,keV} \simeq17$ for $\beta_{\rm 0}=50$, $L_{\rm bol}/L_{\rm 2-10\,keV}\simeq25$ for $\beta_{\rm 0}=100$, and $L_{\rm bol}/L_{\rm 2-10\,keV}\simeq50$ for $\beta_{\rm 0}=200$. This is consistent with the results found in our previous works \citep{2016ApJ...833...35L,2020MNRAS.495.1158C}.
 However, the X-ray photon index obtained by our current model is lower than that in our previous works.
 %%% figure 7 ============

 \section{Model Application}
 The numerical results in the preceding section demonstrate that the X-ray spectrum is significantly affected by the Comptonization of non-thermal electrons. Especially, for a relatively narrow energy range of the non-thermal electrons, the Comptonization in the corona will generate a relatively flat X-ray spectrum with $\Gamma_{\rm 2-10\,keV} < 2.1$. Therefore, we can use this model to reproduce the observed X-ray spectrum with $\Gamma_{\rm 2-10\,keV} < 2.1$.

 \cite{2020MNRAS.495.1158C} applied the refined magnetic-reconnection heating corona model to fit the multi-band spectra of 20 luminous AGNs. In their model, only thermal electrons are included in the corona. In general, the modeling fitting is suitable for the majority of AGN spectra. However, there were four objects (Ton 730, Mrk 290, SBS 1136$+$594, and Mrk 1310) that were not well reproduced because their X-ray spectra are excessively flat. The X-ray spectra of these objects can be fitted by a power-law with photon indices $\Gamma_{\rm 2-10\,keV}$ in the range of $1.53$--$1.64$. The accretion rates of these objects are around $\dot{m}\sim 0.04-0.18$. Besides, PG 1138+222 is also included due to its accretion rate $\dot{m}=0.07$ and its X-ray spectrum fitted by a power-law with photon index $\Gamma_{\rm X}=1.92^{+0.07}_{-0.07}$ \citep{2019MNRAS.487.3884C}. In this section, we apply our model to explain the observational spectra of these AGNs in the optical/UV (UVOT) and X-ray ($2$--$10$\,keV) wavelengths.

 We take the numbers of the black hole mass, the accretion rate, and the outer boundary radius for each object as same as those in \cite{2020MNRAS.495.1158C}, and make sure that the optical/UV spectral data of the AGNs can be well reproduced. Due to the lack of higher-energy data, we only apply our model to produce the hard X-ray emission in the energy band of $2$--$10$\,keV , and set the upper limit of the electron Lorentz factor less than $50$. In general, for each object, we initially set $\beta_{\rm 0}$ to be $50$. If the X-ray luminosity generated by the disc-corona system, which includes only the thermal electrons, is significantly larger than the observed X-ray luminosity, we then increase $\beta_{\rm 0}$. Afterwards, we adjust the parameters of $f_{\rm th}$, $\gamma_{\rm 1}$, $\gamma_{\rm 2}$, and $p$ to reproduce the observed X-ray spectrum. As shown in the subsection \ref{sec:p}, the effect of $p$ on the spectrum is weak when the electron Lorentz factors are in a narrow range. Therefore, we only consider two cases of $p=1.5$ and $p=2.5$ when comparing the observed data.

\begin{figure*}
\begin{center}
\includegraphics[scale=0.4]{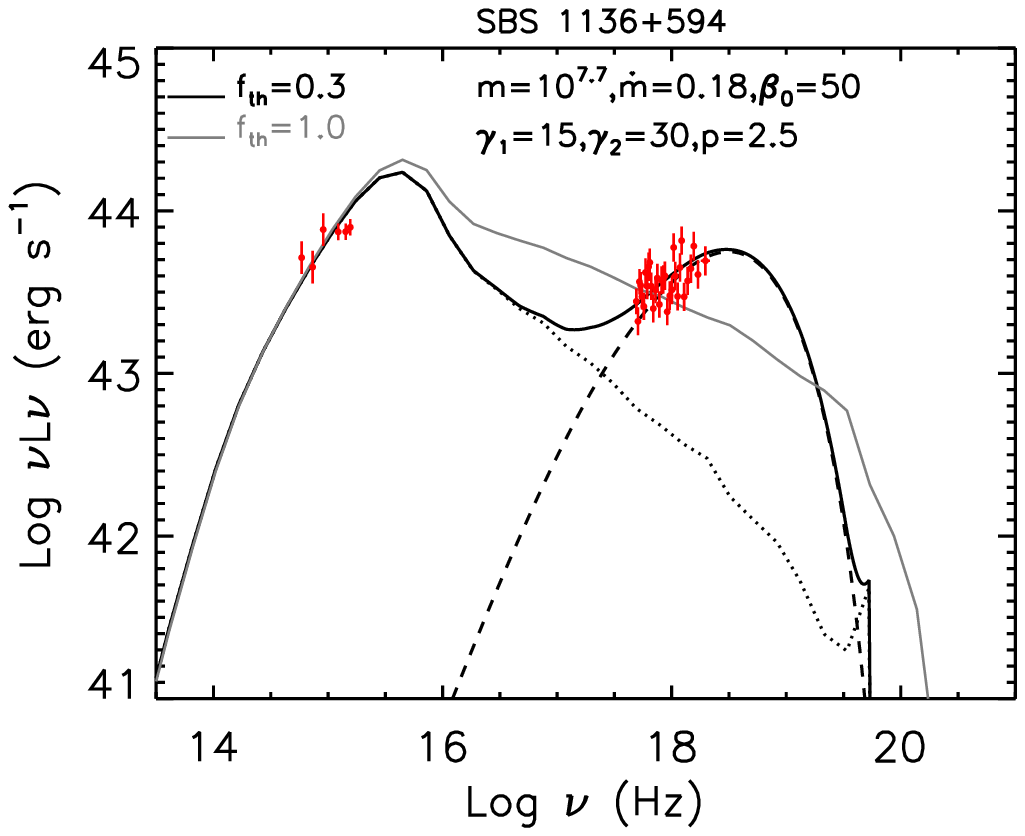}
\includegraphics[scale=0.4]{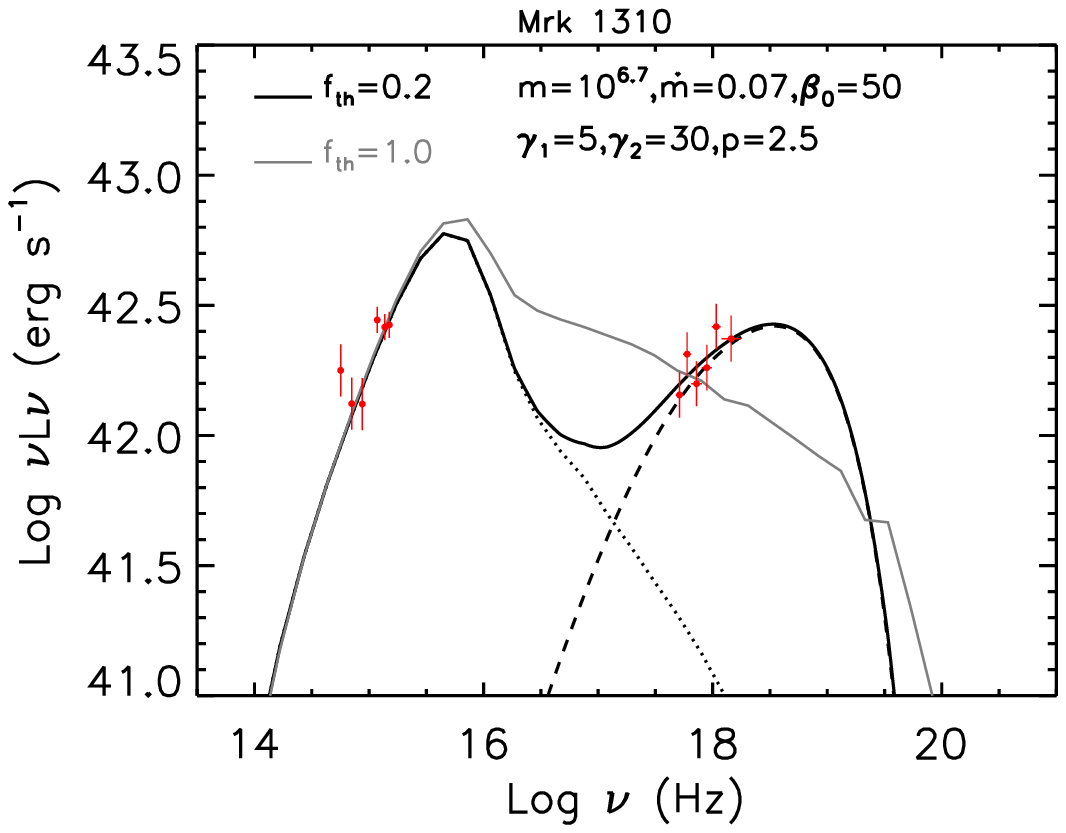}
 \includegraphics[scale=0.4]{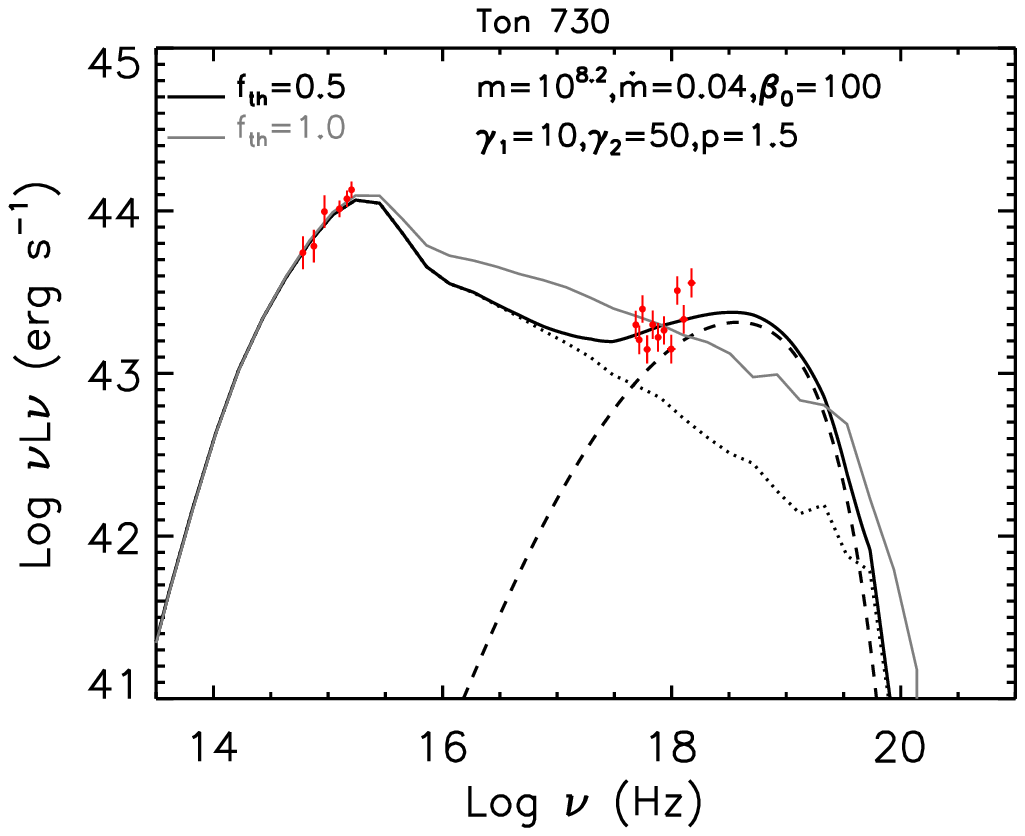}
\includegraphics[scale=0.4]{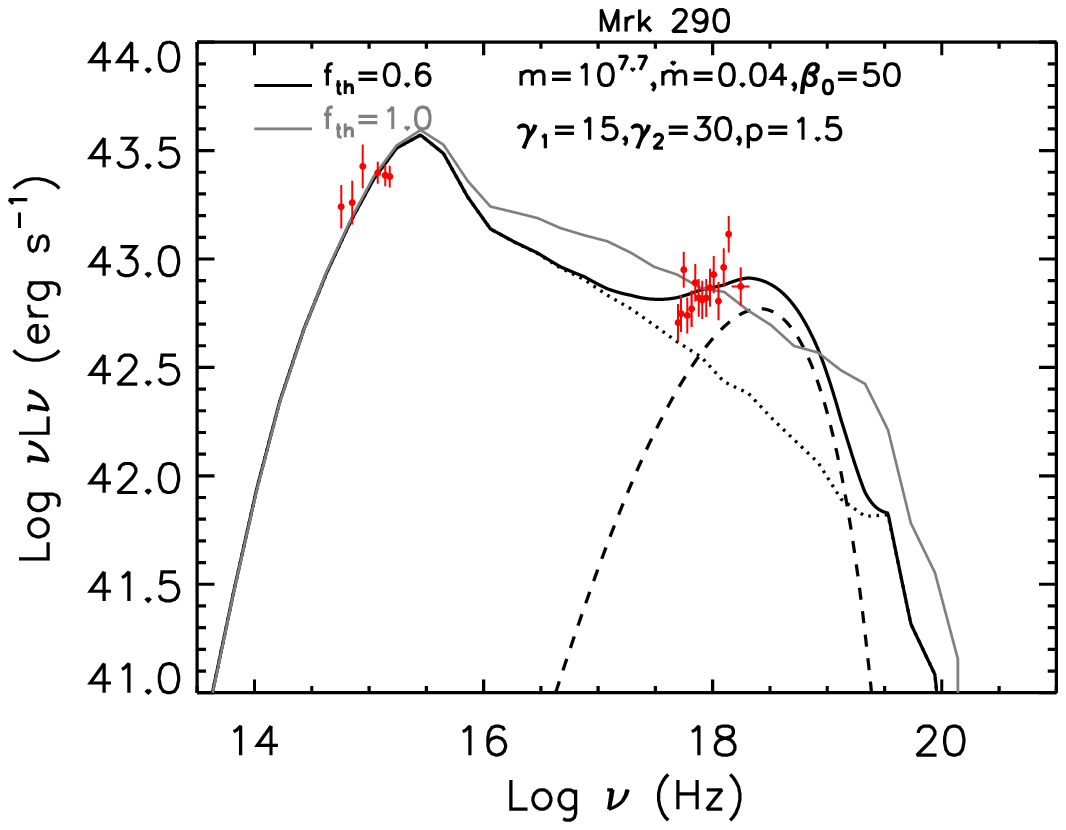}
\includegraphics[scale=0.4]{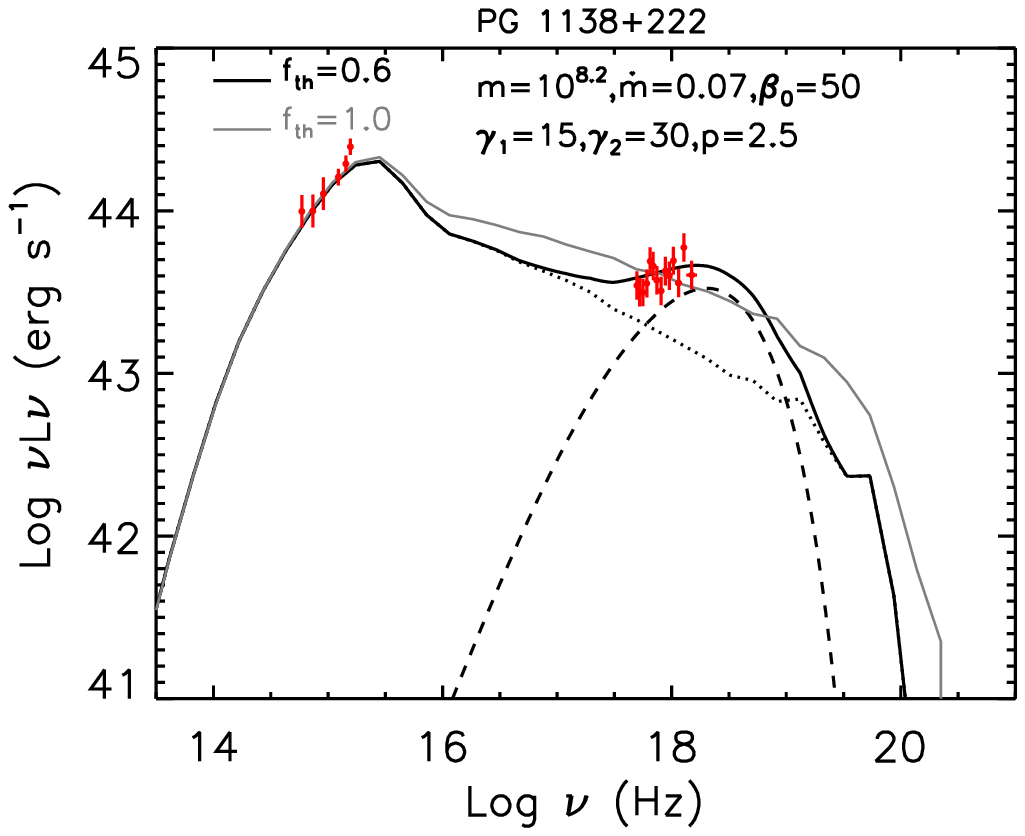}
 \end{center}
\caption{The modeling spectra reproduced by the current model (thick black lines). For comparison, the spectra calculated from the model, which includes only the thermal electrons in the corona, are plotted by the grey lines.}\label{fig:fitting}
\end{figure*}

 The modeling result for each object is plotted by the thick black line in Figure \ref{fig:fitting}. The parameters are noted as $m=M_{\rm BH}/M_{\rm \odot}$ and $\dot{m}=\dot{M}/\dot{M}_{\rm Edd}$. For Ton 730, Mrk 290, and PG 1138+222, their observed luminosity ratios $L_{\rm 2-10\,keV}/L_{\rm UVOT}$ are 0.26, 0.45, and 0.34, respectively. The number of $f_{\rm th}$ is around $0.5$--$0.6$. In other words, more than $40\%$ of the magnetic energy in the corona is allocated to the non-thermal electrons. For SBS 1136+594 and Mrk 1310, their observed luminosity ratio $L_{\rm 2-10\,keV}/L_{\rm UVOT}$ are 0.81 and 1.15, respectively. A smaller $f_{\rm th}$ number ($\sim 0.2-0.3$) is adopted to reproduce their flat X-ray spectra. The grey line in each panel of Figure \ref{fig:fitting} is the spectrum calculated from the model that only includes thermal electrons in the corona. Because the case of pure thermal electrons has the same magnetic field to the case of the hybrid-distributed electrons, the emission reproduced by the pure thermal electrons can be compared to the observational one. However, its photon index is dramatically larger than that obtained from the hybrid-distributed electrons.

In conclusion, the current model containing the hybrid electrons can well reproduce the observed flat X-ray spectra with $\Gamma_{\rm 2-10\,keV}< 2.1$ in luminous AGNs.

%%% figure 8 ============

\section{Discussion} \label{sec:discussion}

%\subsection{Non-thermal Electrons In Magnetic Field}
\subsection{The Energy Distribution of Non-thermal Electrons Accelerated By Magnetic Reconnection}
Particle acceleration in magnetized plasma is an interesting issue in the field of astrophysical research. In some MHD simulation works, the electron energy distribution usually ties up closely with the magnetic field. \cite{2015SSRv..191..545K} reviewed that the non-thermal electrons with a power-law index $p>1$ can be efficiently accelerated by the first-order Fermi acceleration process under the condition of the magnetization parameter $\sigma\equiv B^2/4\pi/(\omega_{n}\rho c^2)>1$,
 where $\omega_{n}=\overline{\gamma}+P/nmc^2$ is the enthalpy per particle,
  $\overline{\gamma}$ is the mean particle Lorentz factor,
   $P$ is the particle pressure, and $n$ is the total particle number density. The power-law index is strongly dependent on the magnetization, with its value decreasing as the $\sigma$ parameter increases \citep{2014PhRvL.113o5005G,2014ApJ...783L..21S,2018MNRAS.473.4840W, 2021ApJ...919..111G, 2021MNRAS.507.5625S}. Especially, \cite{2018MNRAS.473.4840W} systematically investigated the magnetic reconnection process in electron-ion plasmas with $\sigma_{\rm i}=B^2/(4\pi n_{\rm i}m_{\rm i}c^2) <1.0$, where $n_i$ is the proton number density,
  and $m_i$ is the proton mass. This case may be important in accreting black hole corona.
  In this low-magnetization regime, they found that
  the power-law index $p$ of the electron energy spectrum decreases with the increase of
  $\sigma_{\rm i}$, i.e., $p=1.9+0.7/\sqrt{\sigma_{\rm i}}$.
 This result was also confirmed by \cite{2018ApJ...862...80B}.

 In our work, the numerical calculation shows that the X-ray emission mainly originates from the inner region of the disc-corona ($<100$\,Rs). In this region, the magnetic field strength is around $10^2-10^3$\,G and the number density of electrons is around $10^8-10^9~\rm{cm}^{-3}$. According to the disc-evaporation model, the number density of ions is equal to the electron density in a two-temperature corona of pure hydrogen component \citep{2002ApJ...575..117L,2012ApJ...754...81L}. Thus, we estimate the electron magnetization parameter $\sigma_{\rm e}\sim19.4-194$ and the ion magnetization parameter $\sigma_{\rm i}\sim0.01-0.1$. We then obtain the value of $p\sim2.6-1.9$ by the simple estimation method from \cite{2018MNRAS.473.4840W}. Therefore, the value of $p$ adopted in our model seems reasonable.

Some recent studies on magnetic reconnection are focusing on the kinetic lengthscales \citep[e.g.,][]{2017ApJ...850..182L}. Particle acceleration in relativistic turbulence has revealed the energy distribution connection between thermal and non-thermal cases at $\gamma_1\sim 10$ \citep{2018PhRvL.121y5101C}. Further acceleration process in relativistic magnetic reconnection by \cite{2023ApJ...948...19F} seems to support this result.
The particle acceleration driven by magnetic reconnection with the Kelvin-Helmholtz instability was carefully studied by \cite{2021ApJ...907L..44S}.
  When the radiation cooling is taken into account, the reference number of $\gamma_{\rm 1}=75$ was given \citep{2021PhRvL.127y5102C}. \cite{2020MNRAS.493..603Z} obtained the electron energy distribution from the magnetized radiation plasma simulation. The part of the non-thermal electrons is modest shown.
In fact, the initial Lorentz factor of the non-thermal electrons is related to the magnetic field strength in plasmas. \citet{2019ApJ...886..122C} provided $\gamma_{\rm 1}=(1+\sigma_{\rm 0}/2)\gamma_{\rm th}$, where $\sigma_{\rm 0}$ is the magnetization parameter with the range between 2.5 to 80, and $\gamma_{\rm th}=1.6$ represents the initial mean Lorentz factor of Maxwell-J\"{u}ttner thermal electrons. In our work, we take $\gamma_{\rm 1}$ with the range of $5$--$15$ to calculate the X-ray spectrum of the disc-corona system. Our results can be helpful for further constraints on the application of the plasma magnetization and the particle acceleration.

We realize that the magnetization parameter ($\sigma_e$ and $\sigma_i$) and $\gamma_1$ are important for the radiation spectrum. Although we take $f_{\rm{th}}$ as a free parameter, with the comparison to the observational data, we obtain that 40\% magnetic energy is used for the acceleration to the non-thermal electrons. In our framework, the magnetic reconnection with the acceleration process in the disc-corona system should be further investigated in detail in the future. The fraction $f_{\rm{th}}$ obtained in the present work can be one important constraint to the future research.

\subsection{The Synchrotron Emission of Non-thermal Electrons}
 Because our current work focuses on the X-ray emission of hybrid electrons in the corona, we do not consider the synchrotron emission of these hybrid electrons. However, we can estimate the synchrotron emission of non-thermal electrons in the corona as follows.

  For $f_{\rm th}=0.5$, it is shown that the temperature $T_{\rm c}$ is approximately $10^9$\,K and the number density of thermal electrons $N_{\rm e,th}$ is around $10^9$ cm$^{-2}$ in Figure \ref{fig:fth-structure}. We adopt the distribution of non-thermal electrons with $\gamma_{\rm 1}=5$, $\gamma_{\rm 2}=100$, and $p=2.5$. From Equation (\ref{eq:corona1}) and Equation (\ref{eq:corona2}), we can deduce that the number density of non-thermal electrons, $N_{\rm e,pl}$, is approximately to be $10^{-4}N_{\rm e,th}$. The absorption coefficient for the synchrotron, as given by \cite{1985rpa..book.....R}, can be obtained as
 \begin{eqnarray}
 \alpha_{\nu}
 &\simeq&\frac{\sqrt{3}e^3N_{\rm e, pl}C_{\rm 1}}{8\pi m_{\rm e}^2c^2}\left(\frac{3e}{2\pi m_{\rm e}c}\right)^{\frac{p}{2}}\times \nonumber \\
 & &\Gamma\left(\frac{3p+2}{12}\right)\Gamma\left(\frac{3p+22}{12}\right)B^{\frac{p+2}{2}}\nu^{-\frac{p+4}{2}} \nonumber \\
&\simeq& 10^{-6}\left(\frac{B}{10^2\,{\rm G}}\right)^{\frac{9}{4}}\left(\frac{\nu}{10^9\,{\rm Hz}}\right)^{-\frac{13}{4}}\nonumber\\
 \end{eqnarray}
Thus, for the coronal size $\ell_{\rm c}=R\sim10^{13}-10^{15}$\,cm, we obtain a high optical depth $\tau= \alpha_{\nu} \ell_{\rm c}\gg 1$. Consequently, it implies that the synchrotron emission of non-thermal electrons in the corona undergoes significant absorption at GHz frequencies.

\subsection{Polarization}
X-ray polarization is significant to explore the magnetic field properties in the corona and the geometry of the corona. In the corona of AGNs, different polarization features can be produced by the
Compton scattering which is affected by the cross section, the electron components
(thermal, non-thermal, or hybrid), and the geometry of the corona \citep{2010ApJ...712..908S,2017MNRAS.467.2566F,2022MNRAS.510.3674U,2022RAA....22h5011Y}.
The polarization degree in a slab corona is generally larger than that
in a spherical corona, while the polarization degree in a conical corona falls between the two. The effect of non-thermal electrons on the polarization
 properties has been extensively studied by \cite{2017ApJ...850...14B}. Since only about
  15\% of the energy in the corona is carried by the non-thermal electron
   component, the polarization properties in a spherical corona exhibit minimal deviation from those by the fully thermal electrons. However, for the same energy fraction, in the
   optical-thick wedge corona ($\tau=1.9$),
   the polarization degree in the $10$--$100$\,keV energy band of non-thermal
    electrons is larger than that of fully thermalized electrons.

In our work, we find that more than 40\% of the coronal energy is allocated to non-thermal electrons in luminous AGNs with a flat X-ray spectrum.
In the future, we plan to calculate the X-ray polarization by our model
and study the effect caused by the different fractions of non-thermal electrons
in the corona. The results of this future work can be helpful for the observational
 research provided by X-ray polarimetry missions, such as the Imaging X-ray
  Polarimetry Explorer \citep[IXPE,][]{2016SPIE.9905E..17W}, POLAR2 \citep{2020SPIE11444E..2VH}, and the enhanced X-ray Timing and Polarimetry mission \citep[eXTP,][]{2016SPIE.9905E..1QZ}.

\section{Conclusions}
We develop the magnetic-reconnection-heated corona model in which both thermal and non-thermal electrons are included. In the model, the magnetic energy liberated in the corona is cooled through the inverse Compton scattering by the hybrid-distributed electrons. We have found that the parameter, $f_{\rm th}$, describing the fraction of coronal energy allocated to thermal electrons, takes significant effects on the structure and radiation spectrum of the disc-corona system. Additionally, the spectrum strongly depends on both the non-thermal electron energy distribution and the magnetic field. The X-ray luminosity and the X-ray photon index ($\Gamma_{\rm 2-10\,keV}$) depend strongly on both the energy fraction $f_{\rm th}$ and the energy distribution of non-thermal electrons. We apply our model to explain the observed optical/UV and X-ray data of five luminous AGNs with the flat X-ray spectra. We suggest that the disc-corona system can make a flat X-ray spectrum ($\Gamma_{\rm 2-10\,keV}<2.1$) if a large fraction ($>40\%$) of the magnetic energy is allocated to the non-thermal electrons.

\section*{Acknowledgements}
We thank the referee for helpful suggestions and comments. We appreciate Dr. Huaqing Cheng for providing the observed data and Dr. Xiaogu Zhong and Dr. Xiaolin Yang for useful suggestions. This work has the financial support of the National Key R\&D Program of China (2023YFE0101200 and 2021YFA1600402). J.Y. is supported by the Strategic Priority Research Program of Chinese Academy of Sciences, grant No. XDB
41000000, the Natural Science Foundation of Yunnan Province (No. 202201AT070158), the Yunnan Revitalization Talent Support Program Young Talent Project (Notice: The publicity has been completed, but the official document has not yet been issued), the National Natural Science Foundation of China (grants 12133011, 12288102), and the International Centre of Supernovae, Yunnan Key Laboratory (No. 202302AN360001). J.M. is supported by the National Natural Science Foundation of China 11673062, CSST grant CMS-CSST-2021-A06, and the the Yunnan Revitalization Talent Support Program (YunLing Scholar Project). B.F. is supported by the National Natural Science Foundation of China 12073037 and 12333004.

\section*{Data availability}
The data underlying this article will be shared on reasonable request to the corresponding author.
%%%%%%%%%%%%%%%%%%%%%%%%%%%%%%%%%%%%%%%%%%%%%%%%%%

%%%%%%%%%%%%%%%%%%%% REFERENCES %%%%%%%%%%%%%%%%%%

% The best way to enter references is to use BibTeX:

 \bibliographystyle{mnras}
 \bibliography{liujy_et_al} % if your bibtex file is called example.bib

\begin{thebibliography}{}
\makeatletter
\relax
\def\mn@urlcharsother{\let\do\@makeother \do\$\do\&\do\#\do\^\do\_\do\%\do\~}
\def\mn@doi{\begingroup\mn@urlcharsother \@ifnextchar [ {\mn@doi@}
  {\mn@doi@[]}}
\def\mn@doi@[#1]#2{\def\@tempa{#1}\ifx\@tempa\@empty \href
  {http://dx.doi.org/#2} {doi:#2}\else \href {http://dx.doi.org/#2} {#1}\fi
  \endgroup}
\def\mn@eprint#1#2{\mn@eprint@#1:#2::\@nil}
\def\mn@eprint@arXiv#1{\href {http://arxiv.org/abs/#1} {{\tt arXiv:#1}}}
\def\mn@eprint@dblp#1{\href {http://dblp.uni-trier.de/rec/bibtex/#1.xml}
  {dblp:#1}}
\def\mn@eprint@#1:#2:#3:#4\@nil{\def\@tempa {#1}\def\@tempb {#2}\def\@tempc
  {#3}\ifx \@tempc \@empty \let \@tempc \@tempb \let \@tempb \@tempa \fi \ifx
  \@tempb \@empty \def\@tempb {arXiv}\fi \@ifundefined
  {mn@eprint@\@tempb}{\@tempb:\@tempc}{\expandafter \expandafter \csname
  mn@eprint@\@tempb\endcsname \expandafter{\@tempc}}}

\bibitem[\protect\citeauthoryear{{Akylas} \& {Georgantopoulos}}{{Akylas} \&
  {Georgantopoulos}}{2021}]{2021A&A...655A..60A}
{Akylas} A.,  {Georgantopoulos} I.,  2021, \mn@doi [\aap]
  {10.1051/0004-6361/202141186}, \href
  {https://ui.adsabs.harvard.edu/abs/2021A&A...655A..60A} {655, A60}

\bibitem[\protect\citeauthoryear{{Arcodia}, {Merloni}, {Nandra}  \&
  {Ponti}}{{Arcodia} et~al.}{2019}]{2019A&A...628A.135A}
{Arcodia} R.,  {Merloni} A.,  {Nandra} K.,   {Ponti} G.,  2019, \mn@doi [\aap]
  {10.1051/0004-6361/201935874}, \href
  {https://ui.adsabs.harvard.edu/abs/2019A&A...628A.135A} {628, A135}

\bibitem[\protect\citeauthoryear{{Ball}, {Sironi}  \& {{\"O}zel}}{{Ball}
  et~al.}{2018}]{2018ApJ...862...80B}
{Ball} D.,  {Sironi} L.,   {{\"O}zel} F.,  2018, \mn@doi [\apj]
  {10.3847/1538-4357/aac820}, \href
  {https://ui.adsabs.harvard.edu/abs/2018ApJ...862...80B} {862, 80}

\bibitem[\protect\citeauthoryear{{Beheshtipour}, {Krawczynski}  \&
  {Malzac}}{{Beheshtipour} et~al.}{2017}]{2017ApJ...850...14B}
{Beheshtipour} B.,  {Krawczynski} H.,   {Malzac} J.,  2017, \mn@doi [\apj]
  {10.3847/1538-4357/aa906a}, \href
  {https://ui.adsabs.harvard.edu/abs/2017ApJ...850...14B} {850, 14}

\bibitem[\protect\citeauthoryear{{Belmont}, {Malzac}  \& {Marcowith}}{{Belmont}
  et~al.}{2008}]{2008A&A...491..617B}
{Belmont} R.,  {Malzac} J.,   {Marcowith} A.,  2008, \mn@doi [\aap]
  {10.1051/0004-6361:200809982}, \href
  {https://ui.adsabs.harvard.edu/abs/2008A&A...491..617B} {491, 617}

\bibitem[\protect\citeauthoryear{{Beloborodov}}{{Beloborodov}}{2017}]{2017ApJ...850..141B}
{Beloborodov} A.~M.,  2017, \mn@doi [\apj] {10.3847/1538-4357/aa8f4f}, \href
  {https://ui.adsabs.harvard.edu/abs/2017ApJ...850..141B} {850, 141}

\bibitem[\protect\citeauthoryear{{Cao}}{{Cao}}{2009}]{2009MNRAS.394..207C}
{Cao} X.,  2009, \mn@doi [\mnras] {10.1111/j.1365-2966.2008.14347.x}, \href
  {https://ui.adsabs.harvard.edu/abs/2009MNRAS.394..207C} {394, 207}

\bibitem[\protect\citeauthoryear{{Cheng}, {Yuan}, {Liu}, {Breeveld}, {Jin}  \&
  {Liu}}{{Cheng} et~al.}{2019}]{2019MNRAS.487.3884C}
{Cheng} H.,  {Yuan} W.,  {Liu} H.-Y.,  {Breeveld} A.~A.,  {Jin} C.,   {Liu} B.,
   2019, \mn@doi [\mnras] {10.1093/mnras/stz1532}, \href
  {https://ui.adsabs.harvard.edu/abs/2019MNRAS.487.3884C} {487, 3884}

\bibitem[\protect\citeauthoryear{{Cheng}, {Liu}, {Liu}, {Liu}, {Qiao}  \&
  {Yuan}}{{Cheng} et~al.}{2020}]{2020MNRAS.495.1158C}
{Cheng} H.,  {Liu} B.~F.,  {Liu} J.,  {Liu} Z.,  {Qiao} E.,   {Yuan} W.,  2020,
  \mn@doi [\mnras] {10.1093/mnras/staa1250}, \href
  {https://ui.adsabs.harvard.edu/abs/2020MNRAS.495.1158C} {495, 1158}

\bibitem[\protect\citeauthoryear{{Comisso} \& {Sironi}}{{Comisso} \&
  {Sironi}}{2018}]{2018PhRvL.121y5101C}
{Comisso} L.,  {Sironi} L.,  2018, \mn@doi [\prl]
  {10.1103/PhysRevLett.121.255101}, \href
  {https://ui.adsabs.harvard.edu/abs/2018PhRvL.121y5101C} {121, 255101}

\bibitem[\protect\citeauthoryear{{Comisso} \& {Sironi}}{{Comisso} \&
  {Sironi}}{2019}]{2019ApJ...886..122C}
{Comisso} L.,  {Sironi} L.,  2019, \mn@doi [\apj] {10.3847/1538-4357/ab4c33},
  \href {https://ui.adsabs.harvard.edu/abs/2019ApJ...886..122C} {886, 122}

\bibitem[\protect\citeauthoryear{{Comisso} \& {Sironi}}{{Comisso} \&
  {Sironi}}{2021}]{2021PhRvL.127y5102C}
{Comisso} L.,  {Sironi} L.,  2021, \mn@doi [\prl]
  {10.1103/PhysRevLett.127.255102}, \href
  {https://ui.adsabs.harvard.edu/abs/2021PhRvL.127y5102C} {127, 255102}

\bibitem[\protect\citeauthoryear{{Di Matteo}}{{Di
  Matteo}}{1998}]{1998MNRAS.299L..15D}
{Di Matteo} T.,  1998, \mn@doi [\mnras] {10.1046/j.1365-8711.1998.01950.x},
  \href {https://ui.adsabs.harvard.edu/abs/1998MNRAS.299L..15D} {299, L15}

\bibitem[\protect\citeauthoryear{{Done}, {Gierli{\'n}ski}, {Sobolewska}  \&
  {Schurch}}{{Done} et~al.}{2007}]{2007ASPC..373..121D}
{Done} C.,  {Gierli{\'n}ski} M.,  {Sobolewska} M.,   {Schurch} N.,  2007, in
  {Ho} L.~C.,  {Wang} J.~W.,  eds,  Astronomical Society of the Pacific
  Conference Series Vol. 373, The Central Engine of Active Galactic Nuclei.
  p.~121 (\mn@eprint {arXiv} {astro-ph/0703449})

\bibitem[\protect\citeauthoryear{{Fabian}, {Lohfink}, {Belmont}, {Malzac}  \&
  {Coppi}}{{Fabian} et~al.}{2017}]{2017MNRAS.467.2566F}
{Fabian} A.~C.,  {Lohfink} A.,  {Belmont} R.,  {Malzac} J.,   {Coppi} P.,
  2017, \mn@doi [\mnras] {10.1093/mnras/stx221}, \href
  {https://ui.adsabs.harvard.edu/abs/2017MNRAS.467.2566F} {467, 2566}

\bibitem[\protect\citeauthoryear{{French}, {Guo}, {Zhang}  \&
  {Uzdensky}}{{French} et~al.}{2023}]{2023ApJ...948...19F}
{French} O.,  {Guo} F.,  {Zhang} Q.,   {Uzdensky} D.~A.,  2023, \mn@doi [\apj]
  {10.3847/1538-4357/acb7dd}, \href
  {https://ui.adsabs.harvard.edu/abs/2023ApJ...948...19F} {948, 19}

\bibitem[\protect\citeauthoryear{{Gierli{\'n}ski}, {Zdziarski}, {Poutanen},
  {Coppi}, {Ebisawa}  \& {Johnson}}{{Gierli{\'n}ski}
  et~al.}{1999}]{1999MNRAS.309..496G}
{Gierli{\'n}ski} M.,  {Zdziarski} A.~A.,  {Poutanen} J.,  {Coppi} P.~S.,
  {Ebisawa} K.,   {Johnson} W.~N.,  1999, \mn@doi [\mnras]
  {10.1046/j.1365-8711.1999.02875.x}, \href
  {https://ui.adsabs.harvard.edu/abs/1999MNRAS.309..496G} {309, 496}

\bibitem[\protect\citeauthoryear{{Guilbert}, {Fabian}  \& {Rees}}{{Guilbert}
  et~al.}{1983}]{1983MNRAS.205..593G}
{Guilbert} P.~W.,  {Fabian} A.~C.,   {Rees} M.~J.,  1983, \mn@doi [\mnras]
  {10.1093/mnras/205.3.593}, \href
  {https://ui.adsabs.harvard.edu/abs/1983MNRAS.205..593G} {205, 593}

\bibitem[\protect\citeauthoryear{{Guo}, {Li}, {Daughton}  \& {Liu}}{{Guo}
  et~al.}{2014}]{2014PhRvL.113o5005G}
{Guo} F.,  {Li} H.,  {Daughton} W.,   {Liu} Y.-H.,  2014, \mn@doi [\prl]
  {10.1103/PhysRevLett.113.155005}, \href
  {https://ui.adsabs.harvard.edu/abs/2014PhRvL.113o5005G} {113, 155005}

\bibitem[\protect\citeauthoryear{{Guo}, {Li}, {Daughton}, {Li}, {Kilian},
  {Liu}, {Zhang}  \& {Zhang}}{{Guo} et~al.}{2021}]{2021ApJ...919..111G}
{Guo} F.,  {Li} X.,  {Daughton} W.,  {Li} H.,  {Kilian} P.,  {Liu} Y.-H.,
  {Zhang} Q.,   {Zhang} H.,  2021, \mn@doi [\apj] {10.3847/1538-4357/ac0918},
  \href {https://ui.adsabs.harvard.edu/abs/2021ApJ...919..111G} {919, 111}

\bibitem[\protect\citeauthoryear{{Hulsman}}{{Hulsman}}{2020}]{2020SPIE11444E..2VH}
{Hulsman} J.,  2020, in Society of Photo-Optical Instrumentation Engineers
  (SPIE) Conference Series. p. 114442V (\mn@eprint {arXiv} {2101.03084}),
  \mn@doi{10.1117/12.2559374}

\bibitem[\protect\citeauthoryear{{Kagan}, {Sironi}, {Cerutti}  \&
  {Giannios}}{{Kagan} et~al.}{2015}]{2015SSRv..191..545K}
{Kagan} D.,  {Sironi} L.,  {Cerutti} B.,   {Giannios} D.,  2015, \mn@doi [\ssr]
  {10.1007/s11214-014-0132-9}, \href
  {https://ui.adsabs.harvard.edu/abs/2015SSRv..191..545K} {191, 545}

\bibitem[\protect\citeauthoryear{{Kazanas}}{{Kazanas}}{1984}]{1984ApJ...287..112K}
{Kazanas} D.,  1984, \mn@doi [\apj] {10.1086/162668}, \href
  {https://ui.adsabs.harvard.edu/abs/1984ApJ...287..112K} {287, 112}

\bibitem[\protect\citeauthoryear{{Lightman} \& {Zdziarski}}{{Lightman} \&
  {Zdziarski}}{1987}]{1987ApJ...319..643L}
{Lightman} A.~P.,  {Zdziarski} A.~A.,  1987, \mn@doi [\apj] {10.1086/165485},
  \href {https://ui.adsabs.harvard.edu/abs/1987ApJ...319..643L} {319, 643}

\bibitem[\protect\citeauthoryear{{Liu}, {Mineshige}  \& {Shibata}}{{Liu}
  et~al.}{2002a}]{2002ApJ...572L.173L}
{Liu} B.~F.,  {Mineshige} S.,   {Shibata} K.,  2002a, \mn@doi [\apjl]
  {10.1086/341877}, \href
  {https://ui.adsabs.harvard.edu/abs/2002ApJ...572L.173L} {572, L173}

\bibitem[\protect\citeauthoryear{{Liu}, {Mineshige}, {Meyer},
  {Meyer-Hofmeister}  \& {Kawaguchi}}{{Liu}
  et~al.}{2002b}]{2002ApJ...575..117L}
{Liu} B.~F.,  {Mineshige} S.,  {Meyer} F.,  {Meyer-Hofmeister} E.,
  {Kawaguchi} T.,  2002b, \mn@doi [\apj] {10.1086/341138}, \href
  {https://ui.adsabs.harvard.edu/abs/2002ApJ...575..117L} {575, 117}

\bibitem[\protect\citeauthoryear{{Liu}, {Mineshige}  \& {Ohsuga}}{{Liu}
  et~al.}{2003}]{2003ApJ...587..571L}
{Liu} B.~F.,  {Mineshige} S.,   {Ohsuga} K.,  2003, \mn@doi [\apj]
  {10.1086/368282}, \href
  {https://ui.adsabs.harvard.edu/abs/2003ApJ...587..571L} {587, 571}

\bibitem[\protect\citeauthoryear{{Liu}, {Liu}, {Qiao}  \& {Mineshige}}{{Liu}
  et~al.}{2012}]{2012ApJ...754...81L}
{Liu} J.~Y.,  {Liu} B.~F.,  {Qiao} E.~L.,   {Mineshige} S.,  2012, \mn@doi
  [\apj] {10.1088/0004-637X/754/2/81}, \href
  {https://ui.adsabs.harvard.edu/abs/2012ApJ...754...81L} {754, 81}

\bibitem[\protect\citeauthoryear{{Liu}, {Wang}, {Yang}, {Zhu}  \& {Zhou}}{{Liu}
  et~al.}{2014}]{2014ApJ...783..106L}
{Liu} T.,  {Wang} J.-X.,  {Yang} H.,  {Zhu} F.-F.,   {Zhou} Y.-Y.,  2014,
  \mn@doi [\apj] {10.1088/0004-637X/783/2/106}, \href
  {https://ui.adsabs.harvard.edu/abs/2014ApJ...783..106L} {783, 106}

\bibitem[\protect\citeauthoryear{{Liu}, {Taam}, {Qiao}  \& {Yuan}}{{Liu}
  et~al.}{2015}]{2015ApJ...806..223L}
{Liu} B.~F.,  {Taam} R.~E.,  {Qiao} E.,   {Yuan} W.,  2015, \mn@doi [\apj]
  {10.1088/0004-637X/806/2/223}, \href
  {https://ui.adsabs.harvard.edu/abs/2015ApJ...806..223L} {806, 223}

\bibitem[\protect\citeauthoryear{{Liu}, {Qiao}  \& {Liu}}{{Liu}
  et~al.}{2016}]{2016ApJ...833...35L}
{Liu} J.~Y.,  {Qiao} E.~L.,   {Liu} B.~F.,  2016, \mn@doi [\apj]
  {10.3847/1538-4357/833/1/35}, \href
  {https://ui.adsabs.harvard.edu/abs/2016ApJ...833...35L} {833, 35}

\bibitem[\protect\citeauthoryear{{Loureiro} \& {Boldyrev}}{{Loureiro} \&
  {Boldyrev}}{2017}]{2017ApJ...850..182L}
{Loureiro} N.~F.,  {Boldyrev} S.,  2017, \mn@doi [\apj]
  {10.3847/1538-4357/aa9754}, \href
  {https://ui.adsabs.harvard.edu/abs/2017ApJ...850..182L} {850, 182}

\bibitem[\protect\citeauthoryear{{Lubi{\'n}ski} et~al.,}{{Lubi{\'n}ski}
  et~al.}{2016}]{2016MNRAS.458.2454L}
{Lubi{\'n}ski} P.,  et~al., 2016, \mn@doi [\mnras] {10.1093/mnras/stw454},
  \href {https://ui.adsabs.harvard.edu/abs/2016MNRAS.458.2454L} {458, 2454}

\bibitem[\protect\citeauthoryear{{Magdziarz}, {Blaes}, {Zdziarski}, {Johnson}
  \& {Smith}}{{Magdziarz} et~al.}{1998}]{1998MNRAS.301..179M}
{Magdziarz} P.,  {Blaes} O.~M.,  {Zdziarski} A.~A.,  {Johnson} W.~N.,   {Smith}
  D.~A.,  1998, \mn@doi [\mnras] {10.1046/j.1365-8711.1998.02015.x}, \href
  {https://ui.adsabs.harvard.edu/abs/1998MNRAS.301..179M} {301, 179}

\bibitem[\protect\citeauthoryear{{Merloni} \& {Fabian}}{{Merloni} \&
  {Fabian}}{2001}]{2001MNRAS.321..549M}
{Merloni} A.,  {Fabian} A.~C.,  2001, \mn@doi [\mnras]
  {10.1046/j.1365-8711.2001.04060.x}, \href
  {https://ui.adsabs.harvard.edu/abs/2001MNRAS.321..549M} {321, 549}

\bibitem[\protect\citeauthoryear{{Merloni}, {Heinz}  \& {di Matteo}}{{Merloni}
  et~al.}{2003}]{2003MNRAS.345.1057M}
{Merloni} A.,  {Heinz} S.,   {di Matteo} T.,  2003, \mn@doi [\mnras]
  {10.1046/j.1365-2966.2003.07017.x}, \href
  {https://ui.adsabs.harvard.edu/abs/2003MNRAS.345.1057M} {345, 1057}

\bibitem[\protect\citeauthoryear{{Miller} \& {Stone}}{{Miller} \&
  {Stone}}{2000}]{2000ApJ...534..398M}
{Miller} K.~A.,  {Stone} J.~M.,  2000, \mn@doi [\apj] {10.1086/308736}, \href
  {https://ui.adsabs.harvard.edu/abs/2000ApJ...534..398M} {534, 398}

\bibitem[\protect\citeauthoryear{{Poutanen} \& {Vurm}}{{Poutanen} \&
  {Vurm}}{2009}]{2009ApJ...690L..97P}
{Poutanen} J.,  {Vurm} I.,  2009, \mn@doi [\apjl]
  {10.1088/0004-637X/690/2/L97}, \href
  {https://ui.adsabs.harvard.edu/abs/2009ApJ...690L..97P} {690, L97}

\bibitem[\protect\citeauthoryear{{Qiao} \& {Liu}}{{Qiao} \&
  {Liu}}{2015}]{2015MNRAS.448.1099Q}
{Qiao} E.,  {Liu} B.~F.,  2015, \mn@doi [\mnras] {10.1093/mnras/stv085}, \href
  {https://ui.adsabs.harvard.edu/abs/2015MNRAS.448.1099Q} {448, 1099}

\bibitem[\protect\citeauthoryear{{Reeves} \& {Turner}}{{Reeves} \&
  {Turner}}{2000}]{2000MNRAS.316..234R}
{Reeves} J.~N.,  {Turner} M.~J.~L.,  2000, \mn@doi [\mnras]
  {10.1046/j.1365-8711.2000.03510.x}, \href
  {https://ui.adsabs.harvard.edu/abs/2000MNRAS.316..234R} {316, 234}

\bibitem[\protect\citeauthoryear{{Ricci} et~al.,}{{Ricci}
  et~al.}{2017}]{2017ApJS..233...17R}
{Ricci} C.,  et~al., 2017, \mn@doi [\apjs] {10.3847/1538-4365/aa96ad}, \href
  {https://ui.adsabs.harvard.edu/abs/2017ApJS..233...17R} {233, 17}

\bibitem[\protect\citeauthoryear{{Rybicki} \& {Lightman}}{{Rybicki} \&
  {Lightman}}{1985}]{1985rpa..book.....R}
{Rybicki} G.~B.,  {Lightman} A.~P.,  1985, {Radiative processes in
  astrophysics.}.
John Wiley \& Sons

\bibitem[\protect\citeauthoryear{{Schnittman} \& {Krolik}}{{Schnittman} \&
  {Krolik}}{2010}]{2010ApJ...712..908S}
{Schnittman} J.~D.,  {Krolik} J.~H.,  2010, \mn@doi [\apj]
  {10.1088/0004-637X/712/2/908}, \href
  {https://ui.adsabs.harvard.edu/abs/2010ApJ...712..908S} {712, 908}

\bibitem[\protect\citeauthoryear{{Shakura} \& {Sunyaev}}{{Shakura} \&
  {Sunyaev}}{1973}]{1973A&A....24..337S}
{Shakura} N.~I.,  {Sunyaev} R.~A.,  1973, \aap, \href
  {https://ui.adsabs.harvard.edu/abs/1973A&A....24..337S} {24, 337}

\bibitem[\protect\citeauthoryear{{Shemmer}, {Brandt}, {Netzer}, {Maiolino}  \&
  {Kaspi}}{{Shemmer} et~al.}{2006}]{2006ApJ...646L..29S}
{Shemmer} O.,  {Brandt} W.~N.,  {Netzer} H.,  {Maiolino} R.,   {Kaspi} S.,
  2006, \mn@doi [\apjl] {10.1086/506911}, \href
  {https://ui.adsabs.harvard.edu/abs/2006ApJ...646L..29S} {646, L29}

\bibitem[\protect\citeauthoryear{{Sironi} \& {Spitkovsky}}{{Sironi} \&
  {Spitkovsky}}{2014}]{2014ApJ...783L..21S}
{Sironi} L.,  {Spitkovsky} A.,  2014, \mn@doi [\apjl]
  {10.1088/2041-8205/783/1/L21}, \href
  {https://ui.adsabs.harvard.edu/abs/2014ApJ...783L..21S} {783, L21}

\bibitem[\protect\citeauthoryear{{Sironi}, {Rowan}  \& {Narayan}}{{Sironi}
  et~al.}{2021}]{2021ApJ...907L..44S}
{Sironi} L.,  {Rowan} M.~E.,   {Narayan} R.,  2021, \mn@doi [\apjl]
  {10.3847/2041-8213/abd9bc}, \href
  {https://ui.adsabs.harvard.edu/abs/2021ApJ...907L..44S} {907, L44}

\bibitem[\protect\citeauthoryear{{Sridhar}, {Sironi}  \&
  {Beloborodov}}{{Sridhar} et~al.}{2021}]{2021MNRAS.507.5625S}
{Sridhar} N.,  {Sironi} L.,   {Beloborodov} A.~M.,  2021, \mn@doi [\mnras]
  {10.1093/mnras/stab2534}, \href
  {https://ui.adsabs.harvard.edu/abs/2021MNRAS.507.5625S} {507, 5625}

\bibitem[\protect\citeauthoryear{{Sridhar}, {Sironi}  \&
  {Beloborodov}}{{Sridhar} et~al.}{2023}]{2023MNRAS.518.1301S}
{Sridhar} N.,  {Sironi} L.,   {Beloborodov} A.~M.,  2023, \mn@doi [\mnras]
  {10.1093/mnras/stac2730}, \href
  {https://ui.adsabs.harvard.edu/abs/2023MNRAS.518.1301S} {518, 1301}

\bibitem[\protect\citeauthoryear{{Svensson}}{{Svensson}}{1987}]{1987MNRAS.227..403S}
{Svensson} R.,  1987, \mn@doi [\mnras] {10.1093/mnras/227.2.403}, \href
  {https://ui.adsabs.harvard.edu/abs/1987MNRAS.227..403S} {227, 403}

\bibitem[\protect\citeauthoryear{{Svensson} \& {Zdziarski}}{{Svensson} \&
  {Zdziarski}}{1994}]{1994ApJ...436..599S}
{Svensson} R.,  {Zdziarski} A.~A.,  1994, \mn@doi [\apj] {10.1086/174934},
  \href {https://ui.adsabs.harvard.edu/abs/1994ApJ...436..599S} {436, 599}

\bibitem[\protect\citeauthoryear{{Ursini}, {Matt}, {Bianchi}, {Marinucci},
  {Dov{\v{c}}iak}  \& {Zhang}}{{Ursini} et~al.}{2022}]{2022MNRAS.510.3674U}
{Ursini} F.,  {Matt} G.,  {Bianchi} S.,  {Marinucci} A.,  {Dov{\v{c}}iak} M.,
  {Zhang} W.,  2022, \mn@doi [\mnras] {10.1093/mnras/stab3745}, \href
  {https://ui.adsabs.harvard.edu/abs/2022MNRAS.510.3674U} {510, 3674}

\bibitem[\protect\citeauthoryear{{Vasudevan} \& {Fabian}}{{Vasudevan} \&
  {Fabian}}{2007}]{2007MNRAS.381.1235V}
{Vasudevan} R.~V.,  {Fabian} A.~C.,  2007, \mn@doi [\mnras]
  {10.1111/j.1365-2966.2007.12328.x}, \href
  {https://ui.adsabs.harvard.edu/abs/2007MNRAS.381.1235V} {381, 1235}

\bibitem[\protect\citeauthoryear{{Vurm} \& {Poutanen}}{{Vurm} \&
  {Poutanen}}{2009}]{2009ApJ...698..293V}
{Vurm} I.,  {Poutanen} J.,  2009, \mn@doi [\apj] {10.1088/0004-637X/698/1/293},
  \href {https://ui.adsabs.harvard.edu/abs/2009ApJ...698..293V} {698, 293}

\bibitem[\protect\citeauthoryear{{Wang}, {Watarai}  \& {Mineshige}}{{Wang}
  et~al.}{2004}]{2004ApJ...607L.107W}
{Wang} J.-M.,  {Watarai} K.-Y.,   {Mineshige} S.,  2004, \mn@doi [\apjl]
  {10.1086/421906}, \href
  {https://ui.adsabs.harvard.edu/abs/2004ApJ...607L.107W} {607, L107}

\bibitem[\protect\citeauthoryear{{Weisskopf} et~al.,}{{Weisskopf}
  et~al.}{2016}]{2016SPIE.9905E..17W}
{Weisskopf} M.~C.,  et~al., 2016, in {den Herder} J.-W.~A.,  {Takahashi} T.,
  {Bautz} M.,  eds,  Society of Photo-Optical Instrumentation Engineers (SPIE)
  Conference Series Vol. 9905, Space Telescopes and Instrumentation 2016:
  Ultraviolet to Gamma Ray. p. 990517, \mn@doi{10.1117/12.2235240}

\bibitem[\protect\citeauthoryear{{Weng}, {Chen}, {Wang}, {Cai}, {Qiao}  \&
  {Liao}}{{Weng} et~al.}{2020}]{2020MNRAS.491.2576W}
{Weng} S.-S.,  {Chen} Y.,  {Wang} T.-T.,  {Cai} Z.-Y.,  {Qiao} E.,   {Liao}
  N.-H.,  2020, \mn@doi [\mnras] {10.1093/mnras/stz3217}, \href
  {https://ui.adsabs.harvard.edu/abs/2020MNRAS.491.2576W} {491, 2576}

\bibitem[\protect\citeauthoryear{{Werner}, {Uzdensky}, {Begelman}, {Cerutti}
  \& {Nalewajko}}{{Werner} et~al.}{2018}]{2018MNRAS.473.4840W}
{Werner} G.~R.,  {Uzdensky} D.~A.,  {Begelman} M.~C.,  {Cerutti} B.,
  {Nalewajko} K.,  2018, \mn@doi [\mnras] {10.1093/mnras/stx2530}, \href
  {https://ui.adsabs.harvard.edu/abs/2018MNRAS.473.4840W} {473, 4840}

\bibitem[\protect\citeauthoryear{{Yang}, {Wang}  \& {Yang}}{{Yang}
  et~al.}{2022}]{2022RAA....22h5011Y}
{Yang} X.-L.,  {Wang} J.-C.,   {Yang} C.-Y.,  2022, \mn@doi [Research in
  Astronomy and Astrophysics] {10.1088/1674-4527/ac7543}, \href
  {https://ui.adsabs.harvard.edu/abs/2022RAA....22h5011Y} {22, 085011}

\bibitem[\protect\citeauthoryear{{You}, {Cao}  \& {Yuan}}{{You}
  et~al.}{2012}]{2012ApJ...761..109Y}
{You} B.,  {Cao} X.,   {Yuan} Y.-F.,  2012, \mn@doi [\apj]
  {10.1088/0004-637X/761/2/109}, \href
  {https://ui.adsabs.harvard.edu/abs/2012ApJ...761..109Y} {761, 109}

\bibitem[\protect\citeauthoryear{{Yuan}}{{Yuan}}{2003}]{2003ApJ...594L..99Y}
{Yuan} F.,  2003, \mn@doi [\apjl] {10.1086/378666}, \href
  {https://ui.adsabs.harvard.edu/abs/2003ApJ...594L..99Y} {594, L99}

\bibitem[\protect\citeauthoryear{{Zdziarski} \& {Lightman}}{{Zdziarski} \&
  {Lightman}}{1985}]{1985ApJ...294L..79Z}
{Zdziarski} A.~A.,  {Lightman} A.~P.,  1985, \mn@doi [\apjl] {10.1086/184513},
  \href {https://ui.adsabs.harvard.edu/abs/1985ApJ...294L..79Z} {294, L79}

\bibitem[\protect\citeauthoryear{{Zdziarski}, {Poutanen}  \&
  {Johnson}}{{Zdziarski} et~al.}{2000}]{2000ApJ...542..703Z}
{Zdziarski} A.~A.,  {Poutanen} J.,   {Johnson} W.~N.,  2000, \mn@doi [\apj]
  {10.1086/317046}, \href
  {https://ui.adsabs.harvard.edu/abs/2000ApJ...542..703Z} {542, 703}

\bibitem[\protect\citeauthoryear{{Zhang} et~al.,}{{Zhang}
  et~al.}{2016}]{2016SPIE.9905E..1QZ}
{Zhang} S.~N.,  et~al., 2016, in {den Herder} J.-W.~A.,  {Takahashi} T.,
  {Bautz} M.,  eds,  Society of Photo-Optical Instrumentation Engineers (SPIE)
  Conference Series Vol. 9905, Space Telescopes and Instrumentation 2016:
  Ultraviolet to Gamma Ray. p. 99051Q (\mn@eprint {arXiv} {1607.08823}),
  \mn@doi{10.1117/12.2232034}

\bibitem[\protect\citeauthoryear{{Zhdankin}, {Uzdensky}, {Werner}  \&
  {Begelman}}{{Zhdankin} et~al.}{2020}]{2020MNRAS.493..603Z}
{Zhdankin} V.,  {Uzdensky} D.~A.,  {Werner} G.~R.,   {Begelman} M.~C.,  2020,
  \mn@doi [\mnras] {10.1093/mnras/staa284}, \href
  {https://ui.adsabs.harvard.edu/abs/2020MNRAS.493..603Z} {493, 603}

\bibitem[\protect\citeauthoryear{{Zhong} \& {Wang}}{{Zhong} \&
  {Wang}}{2013}]{2013ApJ...773...23Z}
{Zhong} X.,  {Wang} J.,  2013, \mn@doi [\apj] {10.1088/0004-637X/773/1/23},
  \href {https://ui.adsabs.harvard.edu/abs/2013ApJ...773...23Z} {773, 23}

\bibitem[\protect\citeauthoryear{{Zhong} \& {Wang}}{{Zhong} \&
  {Wang}}{2022}]{2022RAA....22c5002Z}
{Zhong} X.-G.,  {Wang} J.-C.,  2022, \mn@doi [Research in Astronomy and
  Astrophysics] {10.1088/1674-4527/ac42c0}, \href
  {https://ui.adsabs.harvard.edu/abs/2022RAA....22c5002Z} {22, 035002}

\bibitem[\protect\citeauthoryear{{Zhou} \& {Zhao}}{{Zhou} \&
  {Zhao}}{2010}]{2010ApJ...720L.206Z}
{Zhou} X.-L.,  {Zhao} Y.-H.,  2010, \mn@doi [\apjl]
  {10.1088/2041-8205/720/2/L206}, \href
  {https://ui.adsabs.harvard.edu/abs/2010ApJ...720L.206Z} {720, L206}

\makeatother
\end{thebibliography}

% Alternatively you could enter them by hand, like this:
% This method is tedious and prone to error if you have lots of references
%\begin{thebibliography}{99}
%\bibitem[\protect\citeauthoryear{Author}{2012}]{Author2012}
%Author A.~N., 2013, Journal of Improbable Astronomy, 1, 1
%\bibitem[\protect\citeauthoryear{Others}{2013}]{Others2013}
%Others S., 2012, Journal of Interesting Stuff, 17, 198
%\end{thebibliography}

%%%%%%%%%%%%%%%%%%%%%%%%%%%%%%%%%%%%%%%%%%%%%%%%%%

%%%%%%%%%%%%%%%%%%%%%%%%%%%%%%%%%%%%%%%%%%%%%%%%%%

% Don't change these lines
\bsp	% typesetting comment
\label{lastpage}
\end{document}